\documentclass[11pt,notitlepage]{article}

\usepackage{amsmath, amssymb}
\usepackage{graphicx}
\usepackage{subfigure}
\usepackage{enumitem}
\usepackage{color}
\usepackage{fancyhdr}
\usepackage{graphicx}
\usepackage{lineno}
\usepackage[margin=0.9in]{geometry}
\usepackage{booktabs}
\usepackage[square,sort,comma,sort&compress,numbers]{natbib}
\usepackage{fleqn}
\usepackage{txfonts}
\usepackage{ulem}
\usepackage{placeins}
\usepackage{hyperref}
\usepackage{dsfont}
\usepackage{authblk}
\usepackage{ulem}

\providecommand{\aMM}{\alpha_{\text{MM}} }

\providecommand{\WMM}{W_{\text{MM}} }

\providecommand{\WXX}{W_{\text{XX}} }
\providecommand{\RMM}{R_{\text{MM}} }

\providecommand{\RXX}{R_{\text{XX}} }
\providecommand{\omegaMM}{\omega_{\text{MM}} }

\providecommand{\Nbin}{N_{\text{bin}} }

\begin{document}
	
	\title{Linking discrete and continuous models of cell birth and migration 
	}
	\author[1]{W.\ Duncan Martinson}
	\author[2]{Alexandria Volkening}
	\author[3]{Markus Schmidtchen}
	\author[4]{Chandrasekhar Venkataraman}
	\author[1]{Jos\'{e} A.\ Carrillo}
	
	\affil[1]{Mathematical Institute, University of Oxford, Oxford, UK}
	\affil[2]{Department of Mathematics, Purdue University, West Lafayette, Indiana, USA}
	\affil[3]{Institute of Scientific Computing, Technische Universit\"at Dresden, Dresden, Germany}
	\affil[4]{Department of Mathematics, University of Sussex, Brighton, UK}
	\maketitle

	\begin{abstract}
		Self-organisation of individuals within large collectives occurs throughout biology. Mathematical models can help elucidate the individual-level mechanisms behind these dynamics, but analytical tractability often comes at the cost of biological intuition. Discrete models provide straightforward interpretations by tracking each individual yet can be computationally expensive. Alternatively, continuous models supply a large-scale perspective by representing the ``effective" dynamics of infinite agents, but their results are often difficult to translate into experimentally relevant insights. We address this challenge by quantitatively linking spatio-temporal dynamics of continuous models and individual-based data in settings with biologically realistic, time-varying cell numbers. Specifically, we introduce and fit scaling parameters in continuous models to account for discrepancies that can arise from low cell numbers and localised interactions. We illustrate our approach on an example motivated by zebrafish-skin pattern formation, in which we create a continuous framework describing the movement and proliferation of a single cell population by upscaling rules from a discrete model. Our resulting continuous models accurately depict ensemble average agent-based solutions when migration or proliferation act alone. Interestingly, the same parameters are not optimal when both processes act simultaneously, highlighting a rich difference in how combining migration and proliferation affects discrete and continuous dynamics.
	\end{abstract}


	\section{Introduction}\label{sec:intro}

	Self-organisation of individual agents is a key feature of life. It occurs ubiquitously throughout the natural world, from the macroscopic example of bird flocking \cite{ME99,d2006self, cucker2007emergent,carrillo2010particle} to the microscopic phenomenon of cell sorting during development \cite{amack2012sorting,burger2018sorting,carrillo2019population,Buttenschon2020,tsai2022differentialadhesion}. The degree to which members of a group coordinate their movement, proliferation, and competition accounts for pattern diversity across biological scales. Alongside experimental approaches, mathematical models can help identify the underlying behaviours that give rise to specific collective dynamics. However, a trade-off often exists between tractability and detail when building models of pattern formation, due in part to the multiscale nature of biological systems. Consequently, better quantitative characterisation of the relationship between analytically tractable models and more biologically representative approaches will improve our understanding of self-organisation throughout nature.
	
\begin{figure}[t!]
		\includegraphics[width=\textwidth]{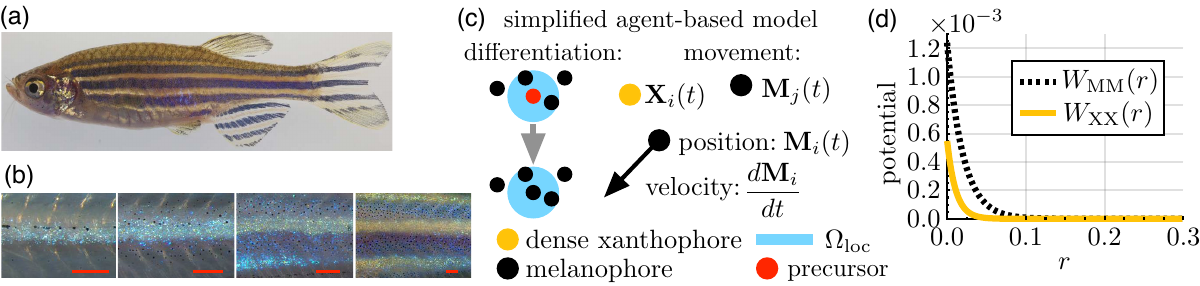}
		\caption{Motivating biological example and model. (a) Wild-type zebrafish feature stripe patterns in their skin. These patterns consist of several types of brightly coloured pigment cells. (b) Over the course of a few months, these cells organise sequentially into stripes and interstripes from the centre of the fish body outward  \cite{frohnhofer2013iridophores}. (c) For the purposes of this manuscript, we focus on a single population of black melanophores or gold dense xanthophores, using a simplified version of the model from \cite{volkening2015}. The agent-based model (ABM) \cite{volkening2015} that motivates our work describes how patterns arise through cell  differentiation, competition, and movement. In our simplified version of the ABM \cite{volkening2015}, we assume new cells appear at randomly selected locations based on short-range activation; this models cell differentiation from uniformly distributed precursors (red position) \cite{volkening2015}, and we also refer to this as ``proliferation" or ``birth" in this paper. (We describe the cell-differentiation rules in the full model \cite{volkening2015} in more detail in Supplementary Fig.~1.) In both our work and the ABM \cite{volkening2015}, cell movement is deterministic and governed by ordinary differential equations (ODEs). (d) These ODEs account for cell--cell repulsion through potential functions, which describe melanophore--melanophore ($\WMM$) and xanthophore--xanthophore ($\WXX$) interactions as a function of their pairwise distance $r$. Red scale bar is $250$ micrometres ($\mu$m) in (b). Image (a) adapted from Fadeev \textit{et al}. \cite{fadeev2015} and licensed under CC-BY $4.0$ (\url{https://creativecommons.org/licenses/by/4.0/}). Image (b) adapted from Frohnh\"{o}fer \textit{et al.} \cite{frohnhofer2013iridophores} and licensed under CC-BY $3.0$ (\url{https://creativecommons.org/licenses/by/3.0/}); published by The Company of Biologists Ltd. \label{fig:biology}}
	\end{figure}

	Here, we help address this open challenge using pigment cell dynamics in zebrafish patterns as a motivation. 
	The zebrafish (\textit{Danio rerio}) is a popular model organism for studying pattern formation, as dark stripes and gold interstripes 
	emerge in its skin during development \cite{frohnhofer2013iridophores,IrionRev2019,Parichy2021,Kondo2021rev}. As we show in Fig.~\ref{fig:biology}, these stripes result from the coordination of interactions among several types of cells, including black melanophores and gold (dense) xanthophores \cite{frohnhofer2013iridophores, nakamasu2009interactions, singh2014proliferation,Gur2020,hamada2014involvement,Inaba}. Experiments that perturb stripes---i.e., by laser ablation \cite{Yamaguchi_Yoshimoto_Kondo_2007,nakamasu2009interactions}---demonstrate how cell--cell signalling and external cues contribute to the creation of alternative motifs such as spots or labyrinths. A rich diversity of mutant patterns, including widened or curvy stripes, also emerge when cell interactions are altered due to genetic mutations \cite{kroll2021simple, hamada2014involvement}.

	Data-driven mathematical models can help uncover the drivers of zebrafish pattern formation and other biological phenomena exhibiting self-organisation by identifying important phase transitions, isolating the effects of specific processes such as cell division, and providing hypotheses that can guide the design of \textit{in vivo} experiments 
	\cite{byrne2010, gompper2020,stillman2023,VolkeningRev}. Different modelling frameworks yield insight at the population or individual level, depending on how they represent members of a group. One modelling approach involves tracking how the position of each individual changes in time. These so-called ``discrete" systems include centre-based models \cite{alert2020,metzcar2019}, cellular automata \cite{deutsch2015cellular,CellularAutomaton17}, cellular Potts models \cite{hirashima2017,rens2019}, and vertex models \cite{alt2017vertex,fletcher2014}. Within the setting of zebrafish patterning, agent-based models (ABMs) have been developed that restrict cells to occupy certain locations ``on-lattice" \cite{Bullara,MorDeutsch,Owen2020} or allow them to roam freely, ``off-lattice", in the domain \cite{caicedo2008silico,volkening2020,volkening2015,volkening2018}. Due to their ability to work on the same length scales as empirical data, ABMs provide an intuitive connection to experiments and allow for detailed predictions about how interactions between agents drive group behaviours. However, ABMs can be prohibitive to simulate when the number of individuals is large, and understanding their long-time behaviour under alternative rules and parameters relies on extensive computation \cite{osborne2017}. 
	
	A second modelling approach uses continuous functions to represent the ``average" density of agents in a collective, with their dynamics governed by a partial differential equation (PDE) in space and time. Continuous models, including reaction-diffusion equations, Boltzmann-like kinetic equations, and integro-differential equations (IDEs), typically cannot resolve individuals and, instead, track the ensemble average (EA) behaviour of a population. However, these models are more amenable to mathematical analysis and more readily provide insight into long-term behaviour than discrete frameworks do  \cite{murray2002mathematical,perthame2006transport}. For example, changes in patterning may arise because of Turing-like instabilities 
	\cite{Turing, maini1997spatial, marciniak2017instability} or due to alterations in physically-based interactions such as cell--cell adhesion \cite{armstrong2006continuum, caicedo2008silico, murakawa2015continuous, carrillo2018zoology, carrillo2018adhesion,carrillo2019population}. In the case of zebrafish patterns, researchers have applied a wide swath of continuous models---including reaction-diffusion equations \cite{Bullara,Gaffney,Konow2021,nakamasu2009interactions,Yamaguchi_Yoshimoto_Kondo_2007} and non-local PDEs \cite{Bloomfield,kondo2017updated,painter,carrillo2019population}---to better understand cell dynamics.
	
	Despite the differences between discrete and continuous approaches, it is possible to establish a mathematical link between these representations in the limit of infinite individuals. This procedure, known as ``coarse-graining", derives differential equations from a given discrete model and yields information about its EA behaviour \cite{giacomin1997phase,painter2002volume,penington2011building,hillen2013,carrillo2014derivation,CK14,bruna22}. For example, the authors in \cite{champagnat2006unifying,champagnat2008individual,champagnat2008individual2} derive logistic IDEs from stochastic processes that describe the birth and death of individuals undergoing Darwinian evolution in the limit of large numbers. Coarse-grained descriptions become inaccurate when relatively few individuals are present, however, as is the case in many biological contexts such as pattern formation in zebrafish. Many approaches also neglect potentially important spatial correlations between cells---caused, for instance, by division or competition---that may play a critical role in pattern dynamics \cite{hansen2006mcdonaldi,lushnikov2008,simpson2009exclusion,simpson2011meanfield,markham2013,wieczorek2023,Konow2021}. While it is possible to go beyond this ``mean-field" setting by deriving continuous models that respect higher-order correlations \cite{binny2015spatial,binny2016collective,middleton2014continuum,bruna2017diffusion,johnston2017new} and hard-core interactions \cite{bruna2012excluded}, these methods still rely on simplifying assumptions that introduce errors between the discrete and continuous frameworks. Controlling these errors in a biological setting is an important objective.

	We help tackle this problem by developing a pipeline to minimise spatio-temporal discrepancies between continuous models and individual-based data in settings with biologically relevant, dynamic cell numbers. The method relies on fitting parameters that effectively dilate the time variable of continuous-model solutions. Our approach can be used to describe biological self-organisation in systems whose macroscopic description can be derived or inferred. We apply our approach to a case study motivated by stripe formation in zebrafish skin, as this allows us to illustrate its utility and interpretablity for experimentally measurable quantities with biologically meaningful spatial and temporal units. In \S\ref{sec:model}, we describe the ABM that we simulate to generate synthetic individual-based data associated with self-organising phenomena. In its full form, the ABM motivating our work \cite{volkening2015} admits pattern formation via non-local rules for cell birth, death, and movement that are inspired by the underlying biology of zebrafish-skin patterns (Fig.~\ref{fig:biology}) \cite{volkening2015,volkening2018}. To focus on the presentation of our pipeline, however, we simplify some biological complexity by reducing this model \cite{volkening2015} to focus on a single cell type---melanophores or dense xanthophores, respectively. (We plan to extend this pipeline to multiple cell types in future work.) We then detail the corresponding continuous descriptions and our method for matching their solutions to EA ABM data even in scenarios with finite and time-dynamic cell numbers. We present our results for black melanophores in \S\ref{sec:results}, and---as a means of demonstrating the generality of our methodology---apply the same approach to dense xanthophores in the Supplementary Information (SI)-\S6.

	\section{Mathematical models and methods} \label{sec:model}

	In \S\ref{sec:modelMove}, we develop our ABM for cell migration and derive its continuous counterpart. Subsequently, in \S\ref{sec:modelBirth}, we introduce our discrete model for cell birth and develop a corresponding continuous IDE model. We present our full models of migration and proliferation in \S\ref{sec:modelFull}. Lastly, we present our approach to estimating scaling parameters in our continuous models from EA ABM data in \S\ref{sec:fitting}. Following previous ABMs \cite{volkening2015,volkening2018} of pattern formation in zebrafish, we assume (1) migration is governed by conservative forces between pigment cells and (2) non-local interactions inform cell birth in a two-dimensional ($2$D) plane \cite{volkening2015,volkening2018,volkening2020}. Throughout this paper, we refer to:
	\begin{align*} 
	\Omega\subset\mathbb{R}^2 &= \text{domain of the simulation with spatial units of millimetres (mm),}\\
	\mathbb{R}^2\ni\textbf{M}_i(t) &= \text{coordinates of the centre of the $i^{\text{th}}$ melanophore at $t$ days in our discrete models,} \\
	\mathbb{N}\ni N_\text{M}(t) &= \text{total number of melanophores present at time $t$ days, and}\\
	\mathbb{R}_{\geq0}\ni M(\textbf{x},t) &= \text{density of melanophores at position $\textbf{x}$ and time $t$ in cells/mm$^2$,}
	\end{align*}  
	with the exception of Fig.~\ref{fig:1dBirth} where we consider a one-dimensional ($1$D) domain; there $M(\mathbf{x},t)$ is the number density of melanophores in cells/mm. Because it appears several times, we define the indicator function $\mathds{1}_{\{\text{condition}\}}$ here, as:
	\begin{align}
	\mathds{1}_{\{\text{condition}\}}(x) &= \begin{cases}
	1, & \text{if $x$ satisfies the rule specified by ``condition"},\\
	0, & \text{otherwise,}
	\end{cases}
	\label{eqn:indicator}
	\end{align}
	where ``condition" depends on the model rule and cell interaction, as we discuss next.
	
	\subsection{Models of migration}\label{sec:modelMove}
	Our ABM for cell movement tracks the position, $\textbf{M}_i(t)$, of each cell, indexed by $i \in \{1,\ldots, N_M(t)\}$, at time $t\geq 0$. The movement of each melanophore depends on its interactions with surrounding melanophores, following an overdamped version of Newton's second law. The forces are assumed to be conservative, i.e., they may be written as the gradient of a potential. This leads to the following system:
	\begin{align} 
	\frac{\text{d}\textbf{M}_i}{\text{d}t} &\approx \textbf{F}^{(i)}_\text{int} = -\sum_{j=1,j\neq i}^{N_\text{M}(t)} \nabla \WMM^c(\textbf{M}_i - \textbf{M}_j).  
	\label{eqn:ABMmoveM}
	\end{align} 
	Here, $\textbf{F}^{(i)}_\text{int}$ is the net force arising from all cell--cell interactions according to the potential $\WMM(\textbf{r})$, where $\textbf{r}$ is the inter-particle displacement. The potential can encode cell--cell repulsion and adhesion, depending on the sign of its gradient along the direction between two cell centres. Many choices for $\WMM(\textbf{r})$ are possible, including harmonic, power-law, Morse, and Lennard-Jones potentials, among others. Here we use exponential potentials given by:
	\begin{align}
	\WMM(\textbf{r}) &= \RMM e^{-|\textbf{r}|/\omegaMM} \ - \ A_\text{MM}e^{-|\textbf{r}|/a_\text{MM}},\label{eqn:Wmm}
	\end{align}
	in accordance with prior ABMs of pattern formation in zebrafish \cite{volkening2018,volkening2015}; see Fig.~\ref{fig:biology}(c) and Table~\ref{table:parameters} for parameter values and their biological interpretations. To model cells communicating through cellular extensions or dendrites \cite{Inaba,hamada2014involvement}, secreted signals \cite{Patterson2014}, or cell--cell contact \cite{Walderich2016}, we assume forces on cells are zero beyond some cut-off distance $d_\text{max}$. We represent this using the notation
	$\nabla \WMM^c (\textbf{r}) =  \nabla  \WMM(\textbf{r}) \, \mathds{1}_{\{|\textbf r|~<~d_\text{max}\}}(\textbf{r})$, and set $N_\text{M}(t) =N_\text{M}$ when there is no cell birth. We remark that the model given by Eqn. \eqref{eqn:ABMmoveM} is deterministic in the sense that it does not include Brownian motion.
	
\begin{figure}[t!]
		\centering
		\includegraphics[width=\textwidth]{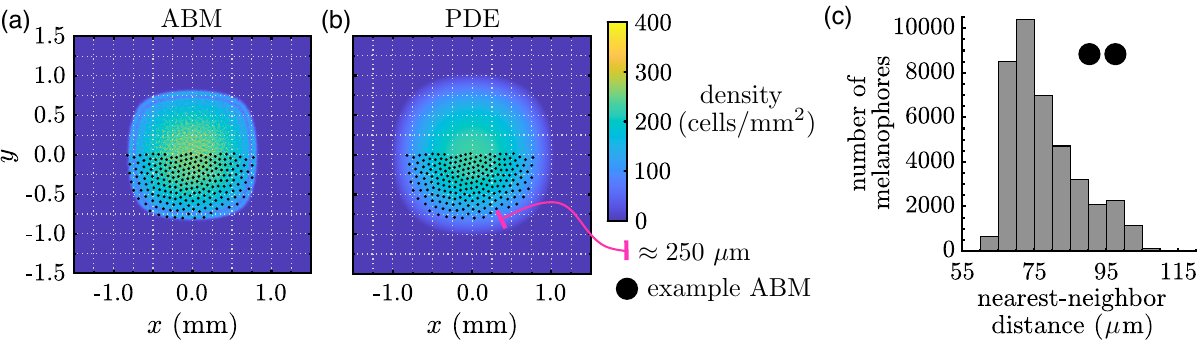}
		
		\caption{The PDE for cell migration does not accurately describe the EA ABM result when its scaling parameter $\aMM$ is set to unity. (a) To compute our EA ABM result, we solve Eqn.~\eqref{eqn:ABMmoveM} using an initial condition of $400$ melanophores placed uniformly at random in a $1$ mm $\times$ $1$ mm square, group cell positions in a $240$ $\times$ $240$ histogram, and average such data over $10^4$ ABM realisations. (b) We compute the corresponding PDE solution by simulating Eqn.~\eqref{eqn:moveM} with $\aMM = 1$ from a uniform density of $400$ cells/mm$^2$ in the same square region. The ABM and PDE solutions use the same potential (given by Eqn.~\eqref{eqn:Wmm} with parameters in Table~\ref{table:parameters}). We overlay an example ABM realisation for comparison; the results demonstrate that the support of the PDE is larger than that of the ABM by about $200$ to $250$ $\mu$m. Because melanophore--melanophore distances have been measured to be roughly $50$ $\mu$m \textit{in vivo} \cite{TakahashiMelDisperse} and stripes are only about $7$--$12$ cells wide \cite{nakamasu2009interactions}, this is a large difference. (c) The distribution of nearest-neighbor distances across $100$ ABM realisations demonstrates that cell--cell separation ranges from roughly $60$ to $100$ $\mu$m. Based on visual inspection of the graphs, nearest-neighbor distances appear inversely proportional to the EA cell density. In (a)--(c), we show results at $t = 150$ days.
		}
		\label{fig:motivation}
	\end{figure}

	The associated continuous model describes the melanophore density, $M(\textbf{x},t)$. Integrating $M(\textbf{x},t)$ over a bounded region yields the total number of melanophores within that area at time~$t$. Following the coarse-graining procedure in \cite{golse,carrillo2014derivation,DF2013}, an outline of which is found in the SI-\S1.1, we obtain the PDE below:
	\begin{align}
	\label{eqn:moveM}
	\frac{\partial M}{\partial t} &= \aMM \nabla\cdot\Big(M \nabla\WMM^c\star M\Big),
	\end{align}
	where the force $\nabla\WMM^c$ is the same as in Eqn.~\eqref{eqn:ABMmoveM} and $\star$ is the convolution operator \cite{volkening2018,volkening2015}. The parameter $\aMM$ in Eqn.~\eqref{eqn:moveM} is not inherent to the coarse-graining procedure; rather, it accounts for possible differences between the discrete and continuous models. Indeed, simulating Eqn.~\eqref{eqn:moveM} with the parameter values listed in Table \ref{table:parameters} and $\aMM = 1$ (the value expected from the mean-field approximation) does not always capture the ABM dynamics; see Fig.~\ref{fig:motivation}. The individual and EA ABM results demonstrate that cells disperse until they are about $55$--$115$ $\mu$m apart at $t = 150$ days. The PDE with $\aMM=1$, however, predicts that cells travel about $250$ $\mu$m further in the same time period. Additionally, the PDE cell density is lower than the EA ABM density near the centre of the domain, implying that cells are more separated there. The continuous solution at earlier times more closely resembles the EA ABM result at $t = 150$ days, however, which suggests that PDE solutions evolve at a faster time scale than that of the discrete model. The parameter $\aMM$ effectively dilates the time variable, such that solutions travel $\aMM$ times more quickly. Thus, a non-unitary value of $\aMM$ is likely to produce a better match between the discrete and continuous solutions. To our knowledge, the value of $\aMM$ cannot be derived \textit{a priori}. Instead, we develop an approach for estimating its value based on ABM data in \S\ref{sec:fitting}, and will pursue an analytic derivation for this parameter in future work.

	\subsection{Models of cell birth}\label{sec:modelBirth}
	
	Our ABM for cell birth consists of stochastic, discrete-time rules which we adapt and simplify from \cite{volkening2015} (motivation for these rules may be found in the SI-\S1.2). Specifically, at each time step (i.e., day) in a simulation we select $N_\text{bir} \in \mathbb{N}$ locations uniformly at random from $\Omega$ and evaluate them synchronously for possible cell birth. Each selected location, \textbf{z}, represents the position of a precursor cell that may differentiate into a melanophore based on the signals that it receives. The conditions for melanophore birth in the ABM \cite{volkening2015} depend on both neighboring melanophores and dense xanthophores, as we show in Supplementary Fig.~1. Since we restrict to one population in this manuscript, we simplify the rules from \cite{volkening2015}; see SI-\S1.2 for details. In particular, a new melanophore emerges at position $\textbf{z}$ according to the rule:
	\begin{align}
	\underbrace{\sum_{i=1}^{N_\text{M}} \mathds{1}_{\{\textbf{M}_i~\in~\Omega_\text{loc}^\textbf{z}\}}(\textbf{M}_i) \ge 1}_\text{short-range activation}~~\text{and}~~\underbrace{\sum_{i=1}^{N_\text{M}}\mathds{1}_{\{\textbf{M}_i~\in~\Omega_\text{loc}^\textbf{z}\}}(\textbf{M}_i) ~<c^+_\text{ABM}}_\text{overcrowding prevention}~~\longrightarrow~~\text{melanophore appears at \textbf{z}},
	\label{eqn:ABMbirth}
	\end{align}
	where $c^+_\text{ABM}\in \mathbb{N}$, and
	\begin{align}
	\Omega_\text{loc}^\textbf{z} &= \text{disk centred at $\textbf{z}$ with radius $d_\text{loc}$}.
    \label{eq:omegaLoc}
	\end{align}
	According to Eqn.~\eqref{eqn:ABMbirth}, new cells appear near existing melanophores until the maximum number of cells---namely, $c^+_\text{ABM}$---in $\Omega_\text{loc}^\textbf{z}$, the interaction region between cells, is reached, see Table \ref{table:parameters} for parameter values. While Eqn.~\eqref{eqn:ABMbirth} is deterministic, stochasticity enters our ABM through our $N_\text{bir}$ randomly selected positions $\{\textbf{z}\}$. Similar stochastic rules can also be used to model cell death, as in \cite{volkening2015,volkening2018}, although we do not consider them here.
	
	We do not know of existing methods for rigorously deriving continuous models of cell birth from off-lattice ABMs with this noise structure. We therefore adopt a phenomenological modelling approach, in which we create a phenomenological continuous model whose governing equations mimic the stochastic interaction rules. We reason that the number density of cells in a continuous model must increase at a constant rate (proportional to $N_\text{bir}$) when continuous versions of the overcrowding and short-range activation restrictions are met, since this occurs at the individual level in the ABM. Furthermore, we represent the density restrictions with an indicator function using the integral of the number density over $\Omega_\text{loc}$ as an argument, since the latter quantity yields the total number of cells within that region. This leads to the continuous model below:
	\begin{align}
	\frac{\partial M}{\partial t}(\textbf{x},t) &= \gamma N_\text{bir}{\ensuremath{\mathds{1}_{\left\{1\leq\int_{\Omega_{\text{loc}}^\textbf{x}}M(\mathbf{y},t)~d\mathbf{y}~<~c^+\right\}}}}(\textbf{x},t),
	\label{eqn:birth}
	\end{align}
	where $c^+$ is the continuous equivalent of the density-limiting parameter $c^+_\text{ABM}$ in Eqn.~\eqref{eqn:ABMbirth}; $N_\text{bir}$ has the same value as in our corresponding; and $\gamma\in\mathbb{R}^+$ is a parameter that effectively dilates the time variable in a similar way as $\aMM$ in the cell-movement model. The units of $\gamma$ must be inversely proportional to those of the domain size in order to make the dimensions of Eqn.~\eqref{eqn:birth} consistent. Its value is unknown, however we can employ a phenomenological argument to determine an expected value by integrating Eqn.~\eqref{eqn:birth} over the whole domain. This yields an upper bound on the number of cells born per unit time of $\gamma N_\text{bir}\vert \Omega\vert$, hence one expects $\gamma\approx  \vert \Omega\vert^{-1}$ to maintain a maximum rate of $N_\text{bir}$ cells born per day as in our ABM. We note, however, that this argument does not take into account possible clustering or other spatial correlations that can occur in the ABM, which may change the values of $\gamma$ and $c^+$ from their expected values. While we could address this by allowing both parameters to depend on the proportion of the domain in which the birth conditions are fulfilled, we leave this extension for a future study and simply estimate uniform values for $\gamma$ and $c^+$ by fitting to EA ABM data, as we do for $\aMM$ in the movement-only model. We overview our approach for estimating the values of $c^+$ and $\gamma$ in \S\ref{sec:fitting}.

	\subsection{Full models of cell movement and birth}\label{sec:modelFull}
	
	We combine our descriptions of cell movement and proliferation to form our full discrete and continuous models. For our full ABM, we move cells according to Eqn.~\eqref{eqn:ABMmoveM} and then introduce new agents based on Eqn.~\eqref{eqn:ABMbirth} at each simulated day; see the SI-\S2 for details. For our continuous model, we combine the terms related to movement and birth, such that the cell density evolves according to:
	\begin{align}
	\frac{\partial M}{\partial t}(\textbf{x},t)  &= \aMM\nabla\cdot\Big(M\nabla W_{\text{MM}}^c\star M\Big) +\gamma N_\text{bir}{\ensuremath{\mathds{1}_{\left\{1~\leq~\int_{\Omega_{\text{loc}}^\textbf{x}}M(\mathbf{y},t)~d\mathbf{y}~<~c^+\right\}}}}(\textbf{x},t),
    \label{eqn:full1}
	\end{align}
	where the parameters $\aMM$, $\gamma$, $N_\text{bir}$, and $c^+$ have the same interpretations as in \S\ref{sec:modelMove} and \S\ref{sec:modelBirth}. Importantly, by assuming that these parameters have the same interpretations, we are assuming that migration and proliferation are additive, so that combining them has no extra influence. Our fitting approach for these parameters, discussed below, allows us to evaluate this choice and better understand the interplay of these two mechanisms in discrete and continuous settings.

	\subsection{Parameter estimation procedure}\label{sec:fitting}

 \begin{table}[t!]
		\centering
		\begin{tabular}{c c p{11.25cm}}
			\toprule
			Parameter & Value & Description and motivation\\
			\toprule
			$\RMM$ & $0.00124$ mm$^2$/day & Strength of melanophore repulsion potential in Eqn.~\eqref{eqn:Wmm}; based on \cite{volkening2020,volkening2015}\\
            \midrule
            $A_\text{MM}$ & $0$ mm$^2$/day & Strength of melanophore adhesion potential in Eqn.~\eqref{eqn:Wmm}; based on \cite{volkening2020,volkening2015}\\
			\midrule
			$\omegaMM$ & $0.02$ mm & Melanophore repulsion interaction range in Eqn.~\eqref{eqn:Wmm}; based on \cite{volkening2020,volkening2015}\\
            \midrule
			$a_\text{MM}$ & $0.012$ mm & Melanophore adhesion interaction range in Eqn.~\eqref{eqn:Wmm}; based on \cite{volkening2020,volkening2015}\\
            \midrule
			$d_\text{max}$ & $0.2$ mm & Maximum cell interaction distance in Eqn.~\eqref{eqn:ABMmoveM}; based on \cite{volkening2020,volkening2015}\\
			\midrule
			$d_\text{loc}$ & $0.075$ mm & Maximum interaction range for cell birth in Eqn.~\eqref{eq:omegaLoc};
            based on \cite{volkening2015} and chosen slightly larger than measurements of cell--cell distances \cite{TakahashiMelDisperse,ParTur256}\\
			\midrule
			$N_\text{bir}$ & Varies & Number of positions selected uniformly at random per day for possible cell proliferation (e.g., differentiation from precursors) in Eqns.~\eqref{eqn:ABMbirth} and \eqref{eqn:birth}\\
			\midrule
			$c^-$ & $1$ cell & Lower bound for the number of cells in a short-range neighborhood for cell proliferation in Eqns.~\eqref{eqn:ABMbirth} and \eqref{eqn:birth} \\
                \midrule
			$c^+$ & $6$ cells & Upper bound for the number of cells in a short-range neighborhood for birth in Eqns.~\eqref{eqn:ABMbirth} and \eqref{eqn:birth}; based on estimations of data \cite{nakamasu2009interactions,TakahashiMelDisperse} in \cite{volkening2015} \\
			\midrule
			$t_\text{final}$ & $150$ or $2,000$ days & Simulation end time ($150$ days in $2$D and $2,000$ days in $1$D)\\ 
			\midrule
			$\Delta t_\text{move}$ & $0.01$ or $0.1$ days & Time step for numerical implementation of Eqns.~\eqref{eqn:ABMmoveM} and (S3)\\
			\midrule
			$\Delta t_\text{bir}$ & $1$ day & Time step for numerical implementation of cell birth in Eqns.~\eqref{eqn:ABMbirth} and \eqref{eqn:birth}\\
			\midrule
			$\Delta t_\text{PDE}$ & $0.05$ days & Time step for numerical implementation of Eqns.~\eqref{eqn:moveM},  \eqref{eqn:full1}, and (S6)\\
			\midrule
			$\Delta t_\text{record}$ & $1$ day & Time step for recording data from model simulations\\
			\midrule
			$N_\text{sim}$ & Varies & Number of ABM realisations for computing EA cell densities\\
                \midrule
			$\Nbin$ & Varies & Spatial discretisation step for solving our continuous models \\
			\midrule
			$N_\text{hist}$ & $\Nbin$ or $30$ voxels & Spatial discretisation step for binning simulations results for comparison \\
			\bottomrule
		\end{tabular}
		\caption{Model and simulation parameters used throughout the paper. We note that $N_\text{bir}~\in \{10, 25, 50, 100, 150, 200, 250\}$ realisations for $2$D simulations and $N_\text{bir} \in \{1,2,3,4,5,6,7,8,9,10\}$ realisations for $1$D simulations. As we discuss in the SI-\S2, $N_\text{sim} = 5\times10^3$ realisations for most EA ABM solutions of the cell-movement model and $N_\text{sim} = 10^3$ for all EA ABM results of the cell birth and combined models. We set $N_\text{hist} = 30$ voxels for EA ABM solutions of the cell birth and combined models, and $N_\text{hist} = N_\text{bin}$ for the cell-movement model. (See figure captions for our $N_\text{bin}$ values.) The value of $\RMM$ (and $\RXX$, see the SI-\S1.3) were reported as repulsion strengths (i.e., $\RMM/\omegaMM$) in \cite{volkening2020}.}
		\label{table:parameters}
	\end{table}
 
\begin{figure}[t]
		\centering
		\includegraphics[width=\textwidth]{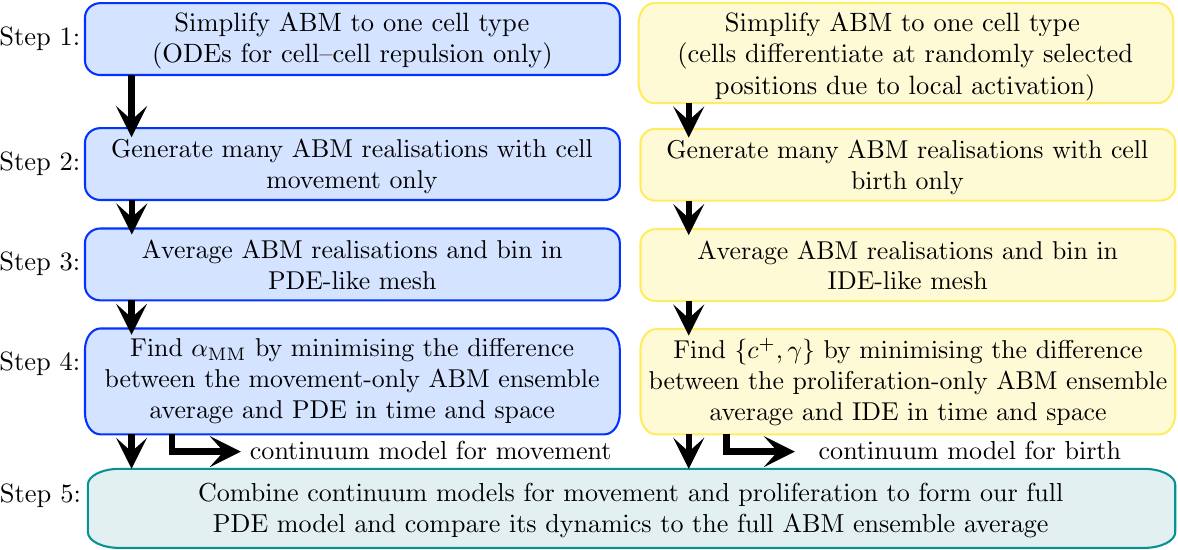}
		\caption{Our modular pipeline for matching the solutions of continuous and discrete models and identifying how cell movement and birth interact in both settings. We first isolate the discrete-model terms from \cite{volkening2015} corresponding to movement (left column) and birth (right column) and simplify them to consider only one cell type. We then produce multiple realisations of our ABMs, sorting the cell locations into a grid of $N_\text{hist}\times N_\text{hist}$ voxels to yield the EA discrete model results. We simulate our continuous model for cell movement (respectively cell birth) and compare it on the same spatial mesh, with values of $\aMM$ (respectively $c^+$ and $\gamma$) obtained from a least squares optimisation approach; see the SI-\S2 for details. Finally, we combine the fitted movement and birth models to produce our full continuous model. While an extension of this pipeline to fit all three parameters simultaneously is straightforward, fitting separately allows us to better understand the effects of cell movement and birth in discrete and continuous frameworks.}
		\label{fig:pipeline}
	\end{figure}

	We identify the values of the parameters---$\aMM$ in Eqn.~\eqref{eqn:moveM} and $\{\gamma,c^+\}$ in Eqn.~\eqref{eqn:birth}---by minimising the sum of squared differences (hereafter referred to as the ``$L^2$ error") between the continuous and EA discrete solutions over time and space. Because we are able to model the isolated processes of cell birth and movement separately, or consider them acting simultaneously, there are two ways of estimating parameters: by fitting all three parameters simultaneously to data from the combined model, or by fitting them in a modular fashion by considering cell movement and birth in isolation from each other. For the remainder of this paper, we adopt a modular approach because it allows us to probe the particular effects of cell movement and birth in detail (Fig.~\ref{fig:pipeline} for an overview), and we present a study of simultaneous estimation in the SI-\S3. In particular, by using our modular parameter values in the combined PDE model, we can investigate their interplay and better understand the additive effects of individual-level mechanisms on the accuracy of continuous models. As we discuss in the SI-\S3, our modular approach may also supply additional information that can improve parameter estimation. For example, we show in the SI-\S3 that $\aMM$ and $\gamma$ are not uniquely identifiable if they are fit only to the combined EA ABM data, whereas Fig.~\ref{fig:convergenceMel}(b) and Fig.~\ref{fig:1dBirth}(f) suggest we can uniquely identify them with a modular approach. We overview our method for parameter estimation below; for parameter values, see Table~\ref{table:parameters}. We refer to the SI-\S2--\S3 for further details about our implementation of the pipeline in addition to alternative choices that could be taken in parameter estimation (such as fitting to earlier times, less refined spatial data, etc.).

	We consider biologically meaningful time scales (i.e., days), length scales (i.e., mm), and cell densities and stress empirical units throughout our results. This choice supports future studies that may treat pattern formation with multiple cell types. Throughout our simulations, we consider a domain of size $3$ mm $\times$ 3 mm (with one $1$D exception in Fig.~\ref{fig:1dBirth}). We implement four initial conditions to extract common features of cell interactions from different geometric scenarios. The first involves a square region of melanophores in the centre of the domain (``Box"), which mimics the symmetry of the domain. For the second initial condition, we place a single stripe of melanophores (``Stripe"), which is motivated by the typical patterning observed in wild-type zebrafish. For the third initial condition, we consider two rectangular regions of melanophores (``Offset rectangles"), which take into account non-standard geometries and the meeting of two disjoint melanophore populations. Finally, the last initial condition we consider involves two melanophore stripes (``Two stripes"), which explores the interactions between two disjoint melanophore populations with biologically realistic sharp fronts. (See Supplementary Fig.~6 for a summary of these initial conditions.) We initialise individual ABM simulations by sampling cell positions uniformly in these regions for each respective initial condition, and initialise our continuous models by setting the cell density uniformly equal to the estimated biological density of $400$ cells/mm$^2$ \cite{volkening2015}. 

	In Step~2 of our pipeline in Fig.~\ref{fig:pipeline}, we solve our discrete models with an explicit approach. Specifically, we solve Eqn.~\eqref{eqn:ABMmoveM} with an explicit forward Euler scheme. To model differentiation from uniformly distributed precursor cells \cite{volkening2015}, we solve the birth-only ABM by selecting $N_\text{bir}$ sites in the domain uniformly at random at a fixed time step (here, 1 day) and  placing a new cell at each position that meets the conditions given by Eqn.~\eqref{eqn:ABMbirth}. (Following the approach in \cite{volkening2015,volkening2018,volkening2020}, we evaluate all $N_\text{bir}$ locations for potential cell proliferation at the same time. This synchronous evaluation means that it is possible, though uncommon, for more than $c^+_\text{ABM}$ cells to be present in a local neighborhood, and the choice of parameters in our model, based on the ABMs \cite{volkening2015,volkening2018,volkening2020}, accounts for this possibility.) We solve our ABM combining migration and birth by simulating Eqn.~\ref{eqn:ABMmoveM} and then implementing cell birth as above, with migration evaluated using a shorter time step than the time step for birth events.

 To compare ABM results directly with the cell density from our continuous models, we obtain an EA distribution by simulating many ABM realisations (here, between $10^3$ and $10^4$ simulations), sorting all the cell locations into a histogram of $N_\text{hist}$ $\times$  $N_\text{hist}$ voxels (or $N_\text{hist}$ $\times$ $1$ voxels in $1$D), and normalising by the number of simulations and the voxel area for each day simulated; see Step~3 of our pipeline in Fig.~\ref{fig:pipeline}. Other ways of relating ABM and PDE results are also possible, for example by introducing Gaussian kernels at each cell location \cite{chiari2022hybrid}. However, as far as the construction of the histogram is concerned, we expect comparable results for particles and localised Gaussian kernels. Furthermore, we show in the SI-\S3 that histogram voxel size used to bin EA ABM data and the final time used to fit the continuous equations only play a minor role in affecting the parameter values we obtain, at least for the cell-movement model.
	
	As part of Step~4 of our pipeline in Fig.~\ref{fig:pipeline}, we need to solve our continuous models, and we do so with explicit approaches. Specifically, we apply a first-order finite volume scheme for the migration model (Eqn.~\eqref{eqn:moveM}), a forward Euler method for the continuous cell birth model (Eqn.~\eqref{eqn:birth}), and a combined finite volume/forward Euler scheme for the full continuous framework (Eqn.~\eqref{eqn:full1}). (More details about the particular time steps used to simulate the discrete and continuous models are in the SI-\S2.)  We simulate the continuous models on an $\Nbin$ $\times$ $\Nbin$ mesh and, to match with EA ABM solutions, record the average cell density at each day on a (possibly coarser) grid of $N_\text{hist}$ $\times$ $N_\text{hist}$ voxels.

	We compute continuous model parameters by minimising either the $L^2$ error between the continuous and EA ABM results across time. Notably, this nonlinear least-squares problem is equivalent to maximum likelihood parameter estimation when the densities produced from the ABM simulations are independent, identically distributed normal random variables with constant variance and mean equal to the continuous solution. The $L^2$ error that we minimise is given by:
    \begin{align}
    	e^2_{L^2} = \lVert M_{\text{cts}} - M_{\text{ABM}} \rVert_{L^2}^2 &=\int_{t=0}^{t=t_\text{final}}\int_{\Omega}\big(M_{\text{cts}}(\mathbf{x},t) - M_{\text{ABM}}(\mathbf{x},t)\big)^2~d\mathbf{x}dt\notag\\
    	&\approx \Delta t_\text{record}\Delta x \Delta y \sum_{n = 0}^{N_T}\sum_{i = 1}^{N_\text{hist}}\sum_{j = 1}^{N_\text{hist}} (M^{(n)}_{\text{cts}, i, j} - M^{(n)}_{\text{ABM}, i, j})^2,
    	\label{eqn:error_appendix}
   	\end{align}
	where $\Delta t_\text{record}$ denotes the time steps at which data are collected; $\Delta x$ and $\Delta y$ are the spatial step sizes of the histogram used to compare the EA ABM and PDE data; $M^{(n)}_{\text{cts}, i, j}$ is the continuous-model solution at time $t_n$ and position $(x_i,y_j)$; and $M^{(n)}_{\text{ABM}, i, j}$ is the corresponding EA ABM result. For the birth-only model we consider fitting to either the $L^2$ error as above or simply the difference in the total cell count of the two data sets (we verify in the SI-\S2 (Supplementary Table 4) that fitting to the $L^2$ error produces similar parameter estimates). When we consider cell birth, we simulate our models with different values of $N_\text{bir}$ and estimate parameters by minimising the sum of the errors across these $N_\text{bir}$ values. We fit parameters related to cell proliferation sequentially---that is, we determine the optimal value for $c^+$ before $\gamma$. We verify in $1$D that sequential and simultaneous estimation does not lead to significant difference in parameter values; see the SI-\S3. 
	
	\section{Results}\label{sec:results}

	We now present our results linking discrete and continuous models of cell migration (\S\ref{sec:modelMove}), birth (\S\ref{sec:modelBirth}), and migration and birth (\S\ref{sec:modelFull}). We first isolate each interaction process, separately identifying the values of $\aMM$ in Eqn.~\eqref{eqn:moveM} and $\{\gamma,c^+\}$ in Eqn.~\eqref{eqn:birth}. As we note in \S\ref{sec:fitting}, this choice allows us to extract the distinct effects of each mechanism. We then determine how this simplification affects the ability of the full continuous framework, given by Eqn.~\eqref{eqn:full1}, 
	to approximate EA ABM solutions. By considering different initial conditions (we discuss the details and motivations for these in the SI-\S4), we demonstrate the robustness of our fitting procedure. Our results show how the time scales of proliferation and movement in our continuous model may depend on numerical implementation and the frequency of stochastic cell birth controlled by $N_\text{bir}$. Moreover, our modular fitting approach highlights important considerations to account for in more general systems where agents are moving and changing in number.

	\subsection{Cell migration}\label{sec:resultsMove}
\begin{figure}[t!]
		\centering
		\includegraphics[width=0.88\textwidth]{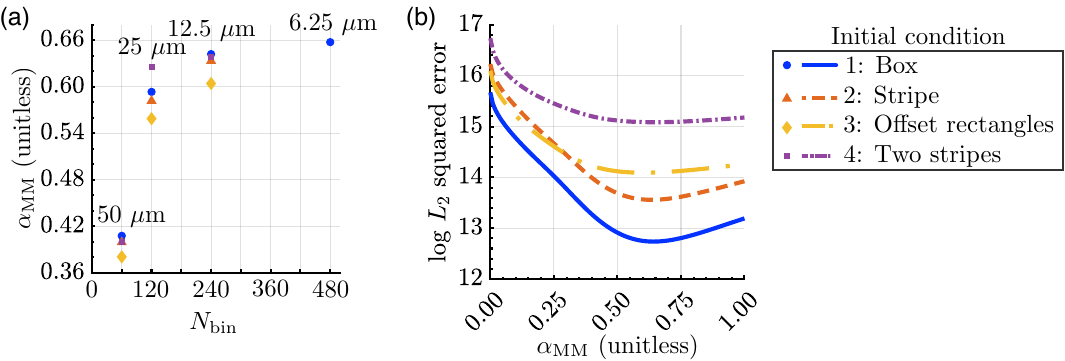}
		\caption{The optimal PDE scaling parameter for movement depends on the mesh resolution but appears to converge. (a) A scatter plot of the numerically optimised value of $\aMM$ from Eqn.~\eqref{eqn:moveM} as a function of the mesh resolution, $\Nbin$, demonstrates that this scaling parameter is correlated with the mesh resolution, but appears to converge at sufficiently high (i.e., $\Nbin\geq240$) detail. (We omit $95$\% confidence intervals because these are so narrow that they are difficult to see.) (b) Plotting the $\log L^2$ error, given by Eqn.~\eqref{eqn:error_appendix}, as a function of $\aMM$ with $\Nbin=240$ for each initial condition suggests that the values in (a) are optimal. As we show in Fig.~\ref{fig:motivation}(c), melanophores are typically separated by more than $60$ $\mu$m in our movement-only ABM simulations---over four times the voxel width in a grid with $\Nbin = 240$. See \S\ref{sec:fitting} and the SI-\S2 for numerical details.
		}
		\label{fig:convergenceMel}
	\end{figure}

	We estimate $\aMM$, the scaling parameter that controls the dynamics of melanophore movement. Fig.~\ref{fig:convergenceMel}(a) presents the values of $\aMM$ that minimise the squared $L^2$ error between the continuous solution of Eqn.~\eqref{eqn:moveM} and EA ABM results for our four initial conditions (see \S\ref{sec:fitting} and the SI-\S2--4). In each case, the optimal value of $\aMM$ is positively correlated with our PDE mesh resolution, i.e., greater values of $\aMM$ are associated with larger $\Nbin=N_\text{hist}$ values. This unitless parameter appears to converge to around $0.60$ to $0.66$ as the mesh resolution increases. There is at most a $2.5$\% relative difference between the values of $\aMM$ that we find when $\Nbin = 240$ versus when $\Nbin = 480$ for our \textit{Box} initial condition. These results suggest that $\aMM$ is independent of the mesh resolution when the latter contains at least $240$ $\times$ $240$ voxels, corresponding to a mesh spacing of $12.5$~$\mu$m. As we show in Fig.~\ref{fig:motivation}(c), melanophores tend to separate by between $60$--$100$ $\mu$m in our ABM results, so this mesh spacing is less than one quarter of the typical distance between agents.
	
	At each mesh resolution in Fig.~\ref{fig:convergenceMel}(a), the estimated optimal value of $\aMM$ does not appear to depend greatly on the initial condition. For example, in the case of a mesh with $\Nbin = N_\text{hist} = 240$, the maximum relative difference between the four parameter values is at most $6.5$\%. This similarity suggests that there is an inherent time scale at which migratory melanophore--melanophore interactions occur. Fig.~\ref{fig:convergenceMel}(b), which presents the $\log L^2$ error for $\Nbin = 240$ as a function of $\aMM$,  further supports this conclusion. Although the errors associated with different initial conditions can vary by an order of magnitude, the minimum value of each (roughly convex) curve appears nearly identical and is located near the values shown in Fig.~\ref{fig:convergenceMel}(a).

	\begin{figure}[t!]
		\centering
		\includegraphics[width=\textwidth]{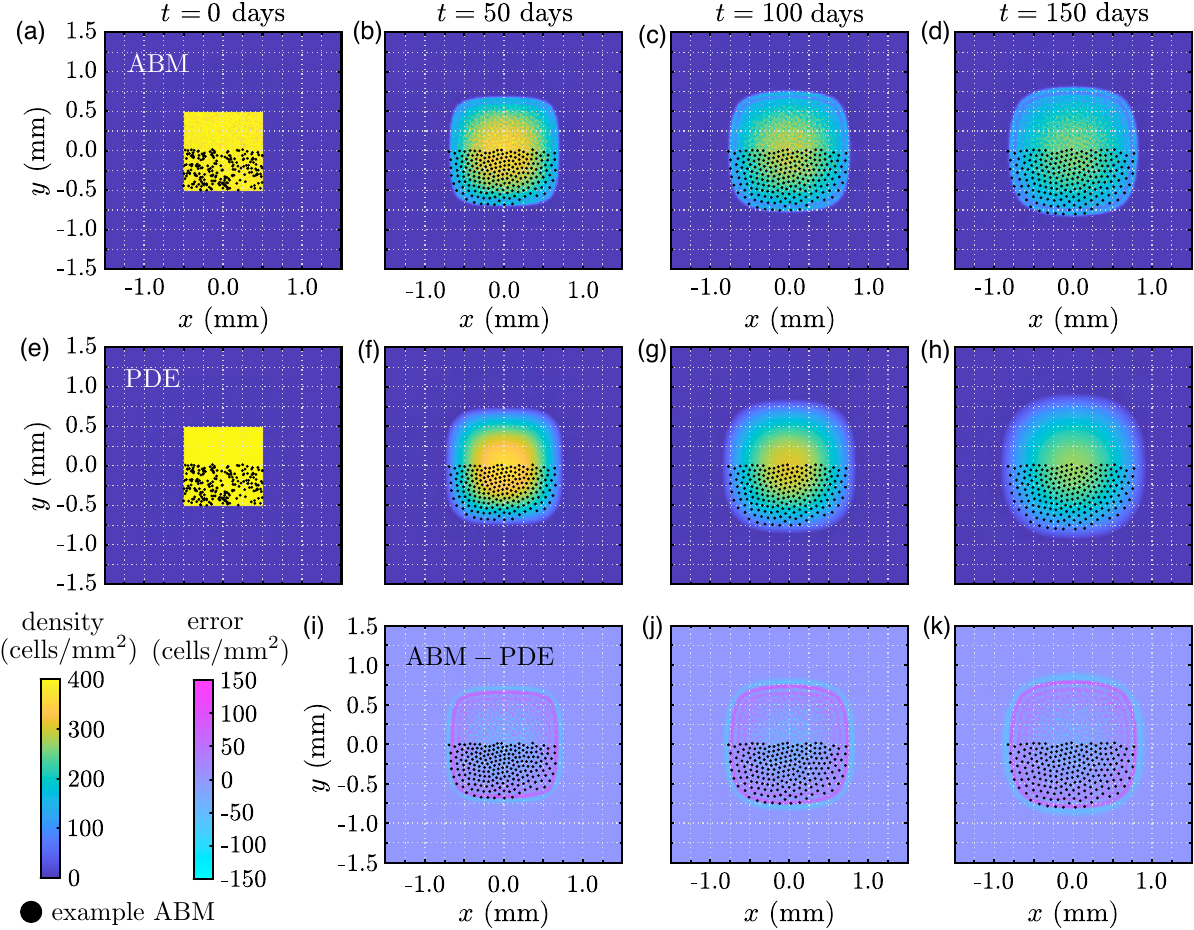}
		\caption{Melanophore movement models with our \textit{Box} initial condition. We present (a)--(d) snapshots of the EA cell density (cells/mm$^2$) across $10^4$ ABM realisations, (e)--(h) the corresponding PDE results using the optimal value of $\aMM$ for a mesh resolution of $\Nbin$ = 240, and (i)--(k) the error between the PDE and EA ABM densities. We overlay cell positions for one example ABM simulation (black points) as a visual guide; we show roughly half of the region occupied by these cells. A difference in cell density of $150$ cells/mm$^2$ in a given voxel corresponds to about $0.0234$ cells for this choice of mesh resolution. We find that the average pointwise errors (over voxels where at least one of the EA ABM or PDE solutions is non-zero) are about $28$ cells/mm$^2$, $17$ cells/mm$^2$, $17$ cells/mm$^2$, and $17$ cells/mm$^2$ at $t = 0, 50, 100$, and $150$ days, respectively. (These values correspond, respectively, to roughly 7\%, 4\%, 4\%, and 4\% of the maximum cell density of $400$ cells/mm$^2$).}
		\label{fig:boxMel}
	\end{figure}

	Fig.~\ref{fig:boxMel} presents snapshots of the EA ABM results across $10^4$ realisations of Eqn.~\eqref{eqn:ABMmoveM} and the optimised PDE solution associated with the \textit{Box} initial condition. The first row shows the expansion in time of the EA ABM support, i.e., the area occupied by the cells, due to melanophore--melanophore repulsion. For more intuition, we superimpose the cell positions from one ABM realisation on our number-density results in this figure and throughout the manuscript. In all cases, we crop out approximately the upper half of cell positions. Visual inspection of cell positions in Fig.~\ref{fig:boxMel} suggests that melanophore--melanophore distances increase near the edge of the collective. Similarly, the speed at which the support expands appears to slow down for the EA ABM result, consistent with melanophores experiencing weaker forces from comparatively distant cells in this region. 
	
	We also observe in Fig.~\ref{fig:boxMel}(a)--(d) that a band of high cell density emerges around the edge of the support which surrounds a ring-like region of low density. These bands may result from the combined effects of cell--cell repulsion and the fine mesh resolution that we use to sort agent positions in the EA solution. Repulsion causes cells at the edge of the collective to travel towards empty regions, while more centrally located agents move more slowly due to the balance of forces from their neighbours. When repulsion separates cells by distances greater than the mesh resolution, we expect regions of low density within the solution support to appear. These oscillatory bands should become less evident when the repulsive potentials in Fig.~\ref{fig:biology}(d) exhibit shallower gradients, as this permits cells to cluster more closely, or when coarser histograms with fewer bins are used to visualise the EA ABM data. As we discuss in the SI-\S6, the forces acting on xanthophores are about an order of magnitude smaller than those for melanophores, and we indeed observe less pronounced bands there. Notably, fitting to EA ABM data on coarser histograms lead to similar parameter estimates; see the SI-\S3 for details.

	We present snapshots of the continuous model, Eqn.~\eqref{eqn:moveM}, under our estimated value of $\aMM$ in Fig.~\ref{fig:boxMel}(e)--(h). This PDE solution captures the dynamics of our example ABM realisation significantly better than the case in Fig.~\ref{fig:motivation}(b), when $\aMM = 1$. However, unlike the EA ABM result,
	the PDE does not exhibit bands of high and low cell density. This discrepancy can be further appreciated in Fig.~\ref{fig:boxMel}(i)--(k), which presents snapshots of the pointwise difference between the PDE and EA ABM solutions. Here positive values indicate that the discrete solution is larger than the continuous one. The lack of bands in the PDE setting is likely because the mean-field assumption used to derive the continuous system is invalid where density is low. We do not expect this discrepancy to be as pronounced in models that include cell birth, as this mechanism increases density; see \S\ref{sec:resultsBoth}. Moreover, the PDE support expands more quickly than that of the ABM. This result is likely due to our choice of error function to fit $\aMM$. Specifically, this parameter is biased towards values that produce accurate approximations in the bulk as these regions have a larger contribution to the $L^2$ norm.
	Since we have already determined the assumptions underlying the continuous model break down in low density regions, however, we choose to fit to the bulk of the cell density and focus on the $L^2$ difference.

	\begin{figure}[t!]
		\centering
		\includegraphics[width=\textwidth]{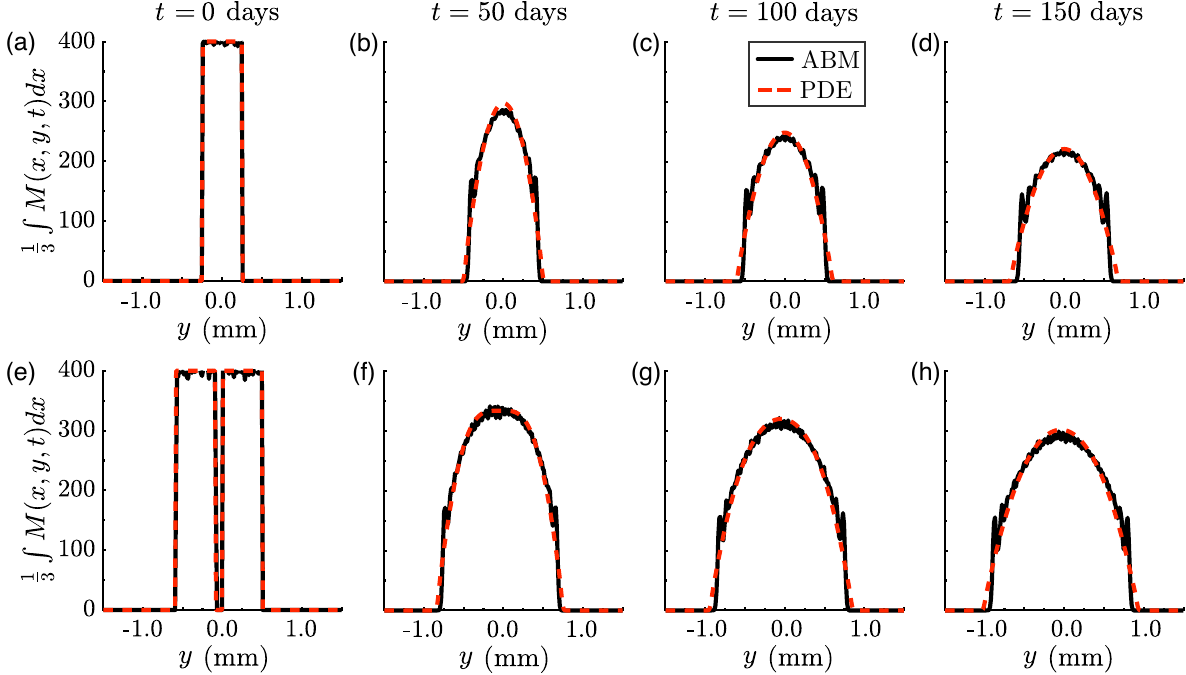}
		\caption{Melanophore movement models with our \textit{Stripe} and \textit{Two stripes} initial conditions. (a)--(d) We present snapshots of the column-averaged cell density (cells/mm$^2$, black solid line), generated from $10^3$ ABM realisations for the \textit{Stripe} case, alongside the corresponding PDE solution (dashed red line) under a mesh resolution of $\Nbin = 240$ and our optimised value of $\aMM$. (e)--(f) Similarly, we show snapshots of the column-averaged density, generated from $10^3$ ABM realisations for the \textit{Two stripes} case, and the corresponding PDE solution. The $2$D solutions are nearly uniform in the $x$-direction (data not shown).}
		\label{fig:stripesMel}
	\end{figure}

	To demonstrate that our observations for the \textit{Box} case are consistent across initial conditions, we compare the EA ABM and PDE dynamics for the \textit{Stripe} and \textit{Two stripes} initial conditions in Fig.~\ref{fig:stripesMel};  Supplementary Fig.~7 presents results for the \textit{Offset rectangles} initial condition. In Fig.~\ref{fig:stripesMel}(a)--(d), the column-averaged PDE solution, i.e., the solution average over the $x$ variable, has a larger support than that of the EA ABM and does not exhibit oscillatory bands. (Comparing column averages is justified because both results are nearly uniform along the $x$-axis.) Nevertheless, the continuous solution closely approximates the EA ABM density, particularly in regions where the latter is high. For example, we find that the average pointwise error (over voxels where at least one of the EA ABM or PDE solution is non-zero) is equal to about $40$ cells/mm$^2$ at $t = 0$ days, $24$ cells/mm$^2$ at $t = 50$ days, $23$ cells/mm$^2$ at $t = 100$ days, and $23$ cells/mm$^2$ at $t = 150$ days (these correspond to roughly 10\%, 6\%, 6\%, and 6\% of the maximum cell density of $400$ cells/mm$^2$, respectively). Both solutions invade empty space in time, and the speed of this travelling wavefront appears to slow as cells become more diffuse. For the \textit{Two stripes} initial condition in Fig.~\ref{fig:stripesMel}(e)--(h), the ABM and PDE predict that cells move into the initially empty space between stripes to approach a characteristic profile also observed in the one-stripe case. The EA ABM model does not appear to form oscillatory bands in the interstripe region, corroborating our hypothesis that these bands are more likely to arise near the edge of the solution support. In this case, the average pointwise error between the continuous and discrete data  
is roughly 18\%, 11\%, 11\% and 11\% of the maximum cell density at $ t = 0, 50, 100,$ and $150$ days, respectively.

	\subsection{Cell birth}\label{sec:resultsBirth}

	\begin{figure}[t!]
		\includegraphics[width=\textwidth]{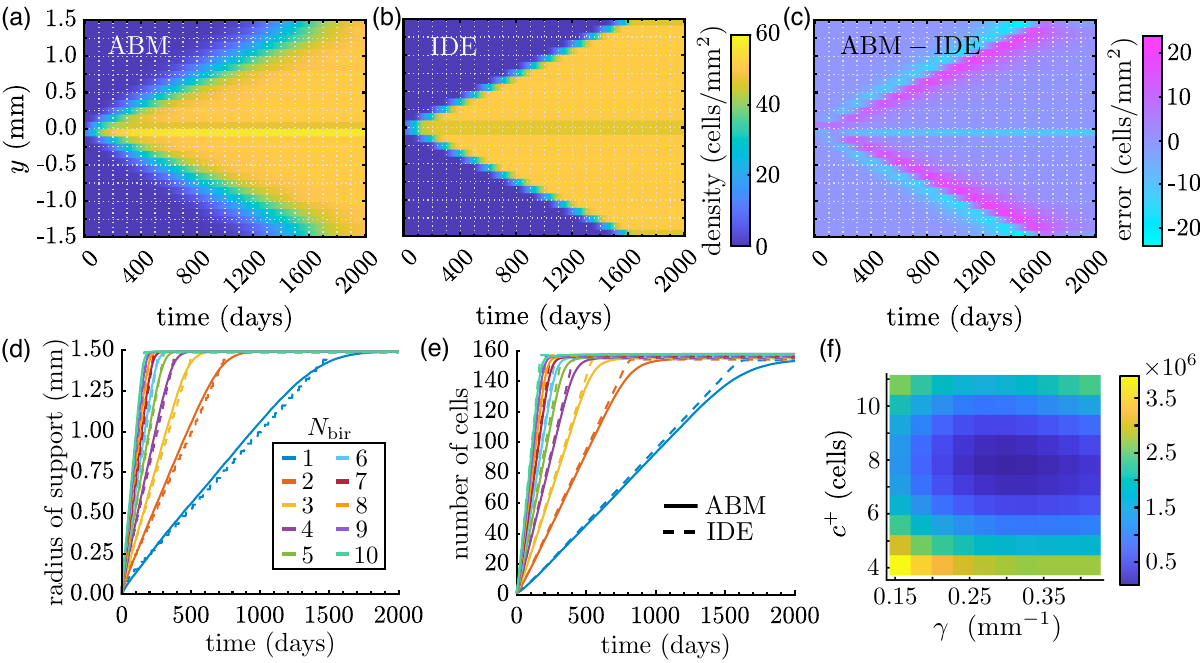}
		\caption{\label{fig:1dBirth}Melanophore birth models with a baseline initial condition of one cell at $y = 0$ in a 1D domain. Results in (a)--(c) are for $N_\text{bir} = 1$ position/day. (a) We compute the EA ABM result by simulating $10^3$ realisations of our ABM birth model, Eqn.~\eqref{eqn:ABMbirth}, and binning cell positions in a histogram with $0.1$~mm-wide voxels (i.e., $N_\text{hist} = 30$). (b) We use a finer mesh resolution to solve our corresponding IDE model (Eqn.~\eqref{eqn:birth}) before transferring results to the same histogram in (a) to perform parameter estimation. Here we show our IDE solution produced with optimal parameter values $c^+ = 7.592$ cells and $\gamma = 0.2822$. (c) The difference between our discrete and continuous results highlights that the ABM support is wider than the PDE support. (d) This is also visible by comparing their mean radii of support in time. To compute the mean radius of support for the ABM at a given time, we find the most distant melanophore from $y=0$ for each simulation and average across these values. (e) Cell mass grows linearly in both models at first, but stochastic effects coupled with our overcrowding condition drive down the growth rate of the ABM as the domain fills with cells. (f) Plotting the squared $L^2$ space-time difference between the discrete and continuous densities, summed over all $N_\text{bir}$ values considered (namely $N_\text{bir} = 1,...,10$), as a function of the density-limiting parameter $c^+$ and birth-rate scaling parameter $\gamma$ highlights its convex shape in $c^+$ and lesser sensitivity to $\gamma$. We compute this $L^2$ difference using a time step of $10$ days here, and our results are based on $10^3$ simulations for each $N_\text{bir}$ value; see the SI-\S3 for parameter values under alternative choices in our estimation process.}
	\end{figure}

	We identify the density-limiting parameter $c^+$ and growth rate $\gamma$ in our IDE model, Eqn.~\eqref{eqn:birth}, by comparing with agent-based data from Eqn.~\eqref{eqn:ABMbirth}. Importantly, the dynamics of discrete-model proliferation, unlike cell migration, involve stochasticity beyond the initial condition. To gain intuition, we thus start with 1D simulations: for each value of $N_\text{bir} \in \{1,2,\dots,10\}$, we compute the EA of $10^3$ ABM realisations from an initial condition in which a single melanophore is placed at the origin in a 1D domain. In Fig.~\ref{fig:1dBirth}(a) we show the EA result for $N_\text{bir} = 1$ and the corresponding IDE model solution with the optimal values of $\gamma$ and $c^+$ in Fig.~\ref{fig:1dBirth}(b). The continuous solution appears to have a smaller radius of support than the EA ABM result at every time point; see Fig.~\ref{fig:1dBirth}(c). This result holds across all $N_\text{bir}$ values in Fig.~\ref{fig:1dBirth}(d). While the IDE predicts a piecewise linear growth of the total number of cells, the corresponding EA ABM result increases linearly before slowly saturating as the domain fills, as we depict in Fig.~\ref{fig:1dBirth}(e). This behaviour likely arises from our overcrowding condition that prevents cell densities from exceeding $c^+$. As the domain fills with cells, it becomes less likely to select a location \textbf{z} that satisfies the overcrowding condition in the ABM. This reduces the population growth rate at later times. In contrast, the IDE model specifies that the support increases by the same amount at each time step until it reaches the domain boundaries. As we discuss in \S\ref{sec:discussion}, capturing discrete model behaviour more accurately at higher cell numbers may require replacing $\gamma$ in our IDE with a density-dependent function.

	Our $1$D simulations provide a baseline case to test our estimation process. As we note in \S\ref{sec:fitting}, we employ a sequential procedure, first fitting $c^+$ with $\gamma = \vert \Omega \vert^{-1}$ and then estimating $\gamma$ with $c^+$ fixed. In the 1D case, this leads to optimal values $c^+ = 7.592$ cells and $\gamma = 0.2822$. If we instead estimate both parameters simultaneously, we find $c^+ = 7.430$ cells and $\gamma = 0.2902$. This is a difference of about $2.1$\% in $c^+$ and $2.8$\% in $\gamma$, suggesting that sequential estimation reduces computational complexity without strongly affecting parameter values. To understand if a coarser discrepancy measure based only on cell numbers at each time is sufficient, we also fit $c^+$ and $\gamma$ by minimising the squared difference in the total cell numbers over time ; see the SI-\S2, Supplementary Table~4 for the resulting parameter values. The corresponding parameter estimates differ from the density-based case by approximately $1.2$\% for and $c^+$ and $\gamma$, suggesting both error measures are reasonable. Both approaches also appear to exhibit similar sensitivity as parameters are varied (compare Figs.\ref{fig:1dBirth}(f) and Supplementary Fig.~2). 
	
	\begin{figure}[t!]
		\includegraphics[width=\textwidth]{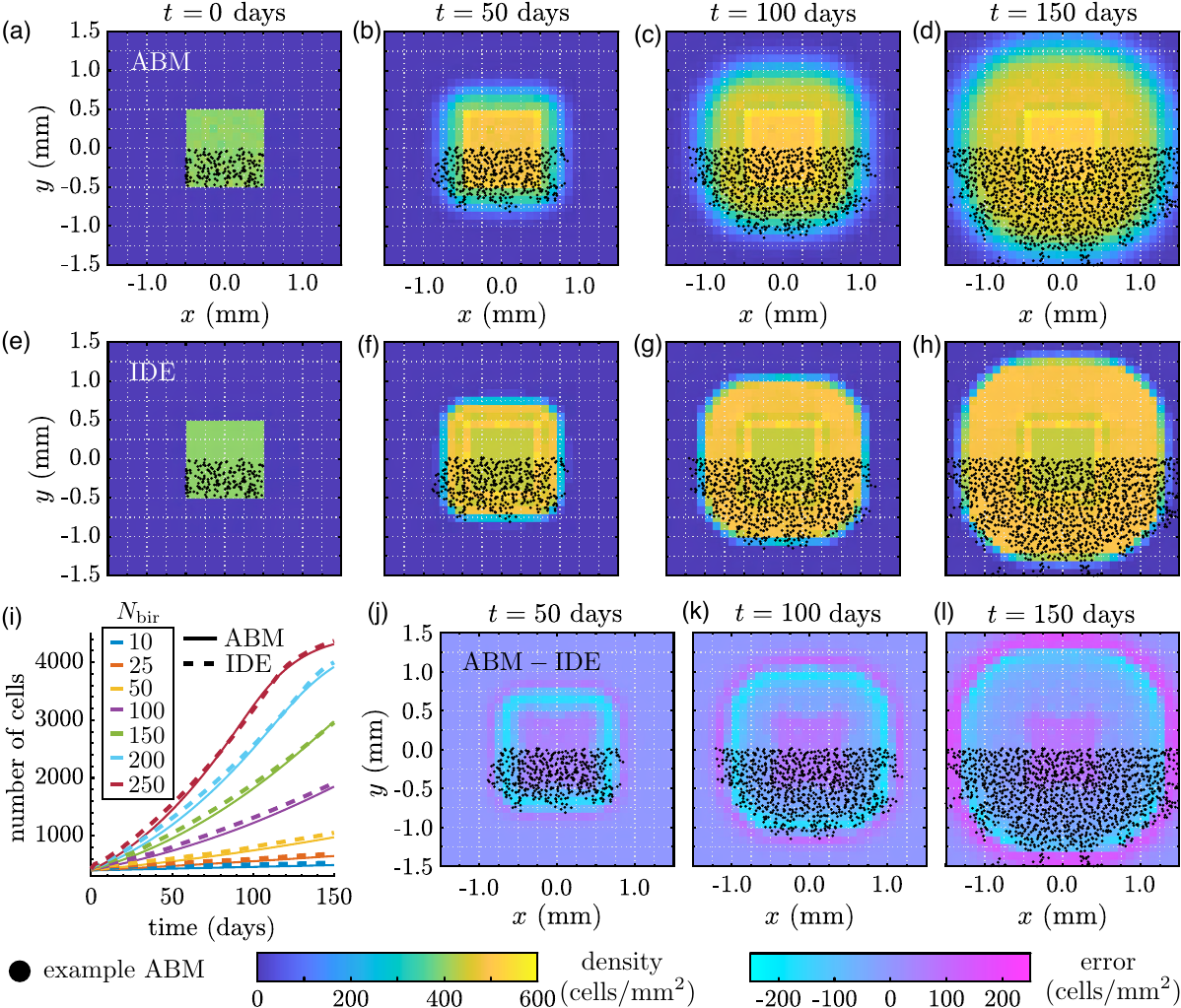}
		\caption{Melanophore proliferation models with our \textit{Box} initial condition. Results in (a)--(h) and (j)--(l) are for $N_\text{bir} = 150$~positions/day. We compute (a)--(d) the EA ABM result across $10^3$ simulations, and (e)--(h) the solution of our IDE model with optimal parameters $c^+ =8.564$ and $\gamma = 0.1274$. (i) Our continuous model captures the mean number of cells in our ABM simulations for different $N_\text{bir}$ values well across time. (j)--(l) As in the 1D case in Fig.~\ref{fig:1dBirth}, the difference between the IDE and EA ABM results demonstrates that the ABM support extends beyond the IDE support. To provide more intuition, we overlay roughly half of the cell positions from an example ABM simulation in (a)--(h) and (j)--(l). See Supplementary Fig.~8 and Supplementary Table~4 for corresponding simulations using our \textit{Offset rectangles} initial condition. \label{fig:boxMelBirth} }
	\end{figure}

	Fig.~\ref{fig:boxMelBirth} and Supplementary Fig.~8, respectively, show that proliferation in 2D broadens the solution support from the \textit{Box} and \textit{Offset rectangles} initial conditions over time, and the IDE model accurately captures the total cell mass of the ABM system for all $N_\text{bir}$ values considered. Our estimated optimal values of $c^+$ and $\gamma$ for these two initial conditions differ by about $2.5$\% and $0.31$\%, respectively, suggesting that our estimation procedure is robust to the initial condition. We also highlight that a region of higher density forms at the edge of the initial condition's support for both the ABM and IDE in Fig.~\ref{fig:boxMelBirth}(a)--(h). Indeed, if $\textbf{z}$ is near the support boundary, $\Omega_\text{loc}^{\textbf{z}}$ covers only a fraction of the occupied domain, thereby meeting both conditions for birth. Conversely, the cell density at the centre of the domain is comparatively low throughout time because the total number of cells contained within disks of size $\lvert\Omega_\text{loc}\rvert$ is already close to the threshold $c^+$. Interestingly, as in the 1D case with only proliferation, the ABM EA support is larger than that of the IDE solution, the reverse of the behaviour that we observed for cell migration in Fig.~\ref{fig:boxMel}.

	\subsection{Cell movement and proliferation}\label{sec:resultsBoth}

	\begin{figure}[t!]
		\includegraphics[width=\textwidth]{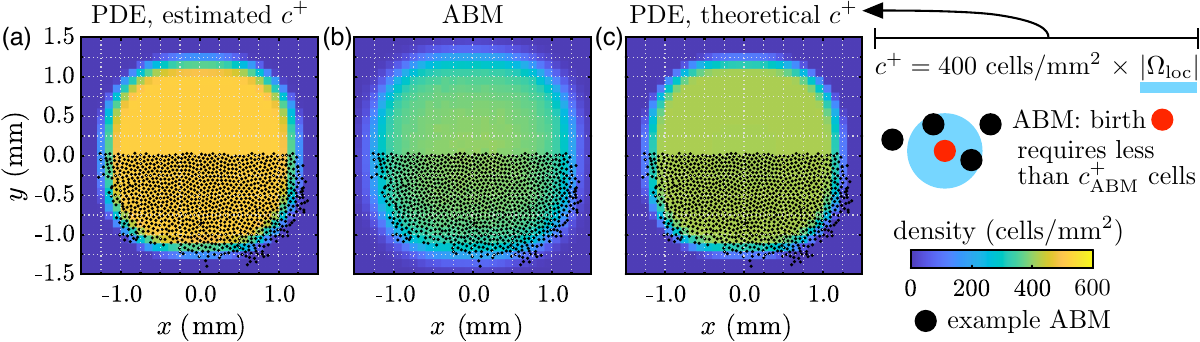}
		\caption{A modular approach to fitting parameters for cell movement and birth does not account for the interplay between these two mechanisms. We show results in (a)--(c) at $t=100$ days for $N_\text{bir} =150$ positions/day. (a) The solution of our full PDE model (Eqn.~\eqref{eqn:full1}) with the values of $\aMM$ and $\{c^+,\gamma\}$ that we fit based on ABM simulations of cell movement and birth, respectively, captures the support of the ABM EA result, but not its density. (b) In comparison, the density for the full discrete model is roughly $400$ cells/mm$^2$. (c) By integrating this density, which is based on empirical estimates of melanophore--melanophore distances \cite{TakahashiMelDisperse,volkening2015}, over an $\Omega_\text{loc}$-region, we find that $c^+ \approx 7.0686$~cells. With this value of $c^+$, alongside the values of $\aMM$ and $\gamma$ that we estimated for migration and birth individually, our PDE produces cell densities that more accurately represent the ABM dynamics.\label{fig:theoreticalC}}
	\end{figure}
	
	To obtain a full continuous 
	model, we may substitute our estimated values of the migration scaling parameter $\aMM$, density-limiting parameter $c^+$, and birth-rate scaling parameter $\gamma$ into Eqn.~\eqref{eqn:full1}. However, comparing this model to the dynamics of our full ABM shows that migration and proliferation have interwoven effects. To illustrate this phenomenon, we present a PDE solution with our optimal values of $\aMM$, $c^+$, and $\gamma$ from \S\ref{sec:resultsMove} and \ref{sec:resultsBirth} at $t= 70$ days in Fig.~\ref{fig:theoreticalC}(a). We observe that this PDE model produces a significantly higher cell density than its discrete counterpart in Fig.~\ref{fig:theoreticalC}(b). This discrepancy occurs regardless of the value of $N_\text{bir}$, which influences the speed of cell birth. Related to this, we notice that the long-time cell density in our ABM results is much lower when both mechanisms operate simultaneously than it is when only birth occurs; compare Fig.~\ref{fig:boxMelBirth}(d) and Fig.~\ref{fig:theoreticalC}(b). On the other hand, the inclusion of movement does not influence the long-time density of the continuous model solution; see the colourbar in Fig.~\ref{fig:boxMelBirth}(h) in comparison to the one in Fig.~\ref{fig:theoreticalC}.
	Although we do not furnish these observations with an analytical explanation here, they demonstrate an interesting difference in how ``adding" mechanisms or terms impact PDE and ABM dynamics.

\begin{figure}[t!]
		\includegraphics[width=\textwidth]{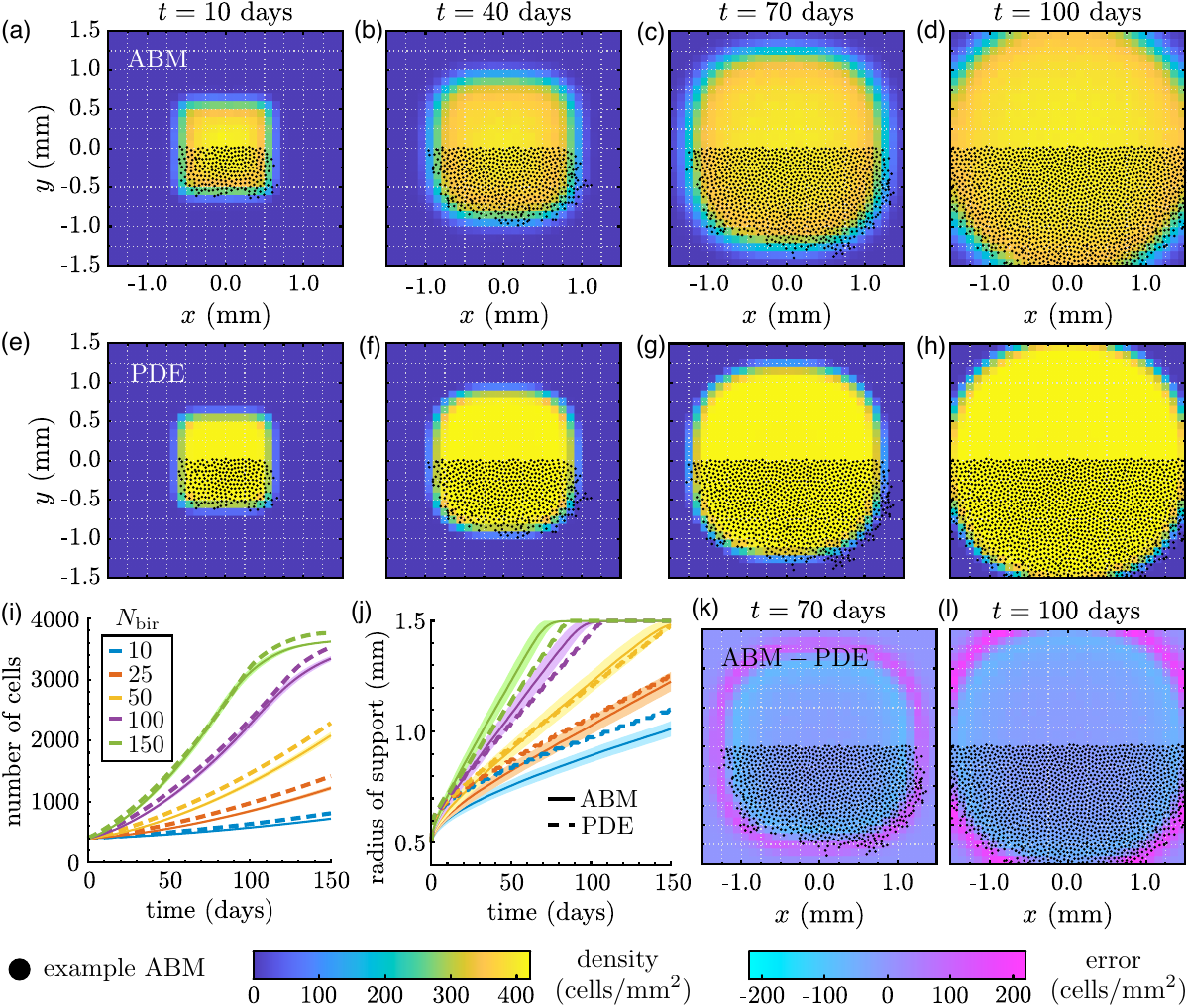}
		\caption{Melanophore movement and birth models with our \textit{Box} initial condition. Results in (a)--(h) and (j)--(l) are for $N_\text{bir} = 150$ positions/day. We (a)--(d) compute the EA ABM result using $10^3$ simulations, and (e)--(h) generate the PDE solution of Eqn.~\eqref{eqn:full1} with $c^+ = 7.0686$ cells and the values of $\aMM$ and $\gamma$ that we estimated in \S\ref{sec:resultsMove} and \S\ref{sec:resultsBirth}, respectively. (i) The time evolution of the PDE cell mass agrees well with the mean number of cells for the ABM under different $N_\text{bir}$ values. (j) Depending on the time scales of migration and birth, the approximate PDE radius of support overtakes or trails the corresponding EA ABM result. We compute the radius of support for each ABM realisation by finding the most distant cell from the origin at each time step; we then average these values across our simulations. In the PDE case, we find the furthest voxel with non-zero density from the origin based on the $L^\infty$ distance, after setting the density to zero if it is below single-digit precision of $10^{-7}$. (k)--(l) We show the difference between the PDE and EA ABM solutions from (a)--(h) at two sample times. We overlay cell positions form one ABM simulation to illustrate how the continuous and discrete solutions are related. In (i) and (j), shaded regions denote plus or minus one standard deviation of the EA ABM solution. \label{fig:boxMelFull}}
	\end{figure}

\begin{figure}[t!]
		\includegraphics[width=\textwidth]{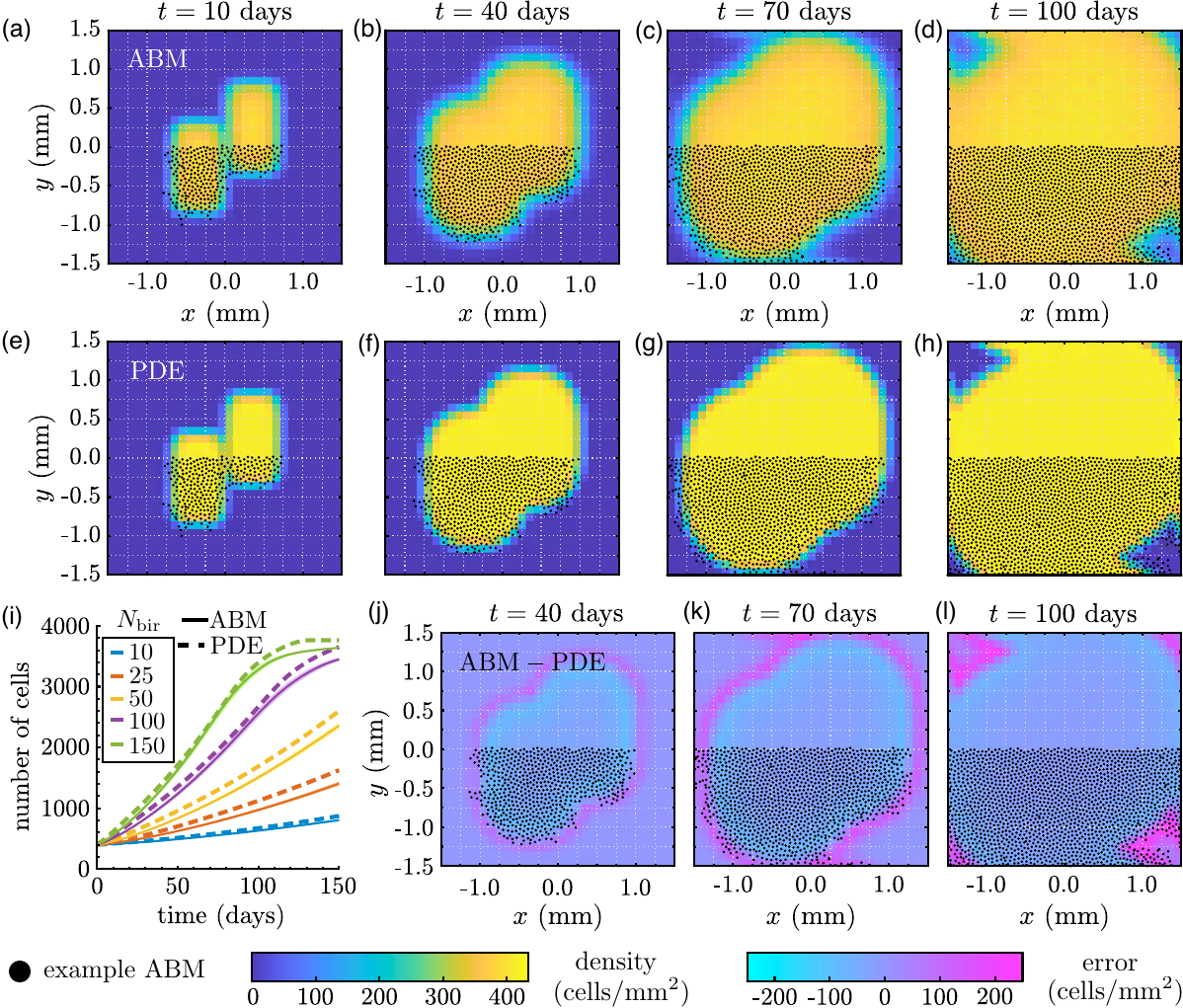}
		\caption{Melanophore movement and birth models with our \textit{Offset rectangles} initial condition to verify that our fitting procedure is robust in non-standard geometries. Results in (a)--(h) and (j)--(l) are for $N_\text{bir} = 150$ positions/day. As in Fig.~\ref{fig:boxMelFull}, we show (a)--(d) the EA ABM result across $10^3$ simulations; (e)--(h) the corresponding PDE solution with $c^+ = 7.0686$ cells and the values of $\aMM$ and $\gamma$ that we estimated in \S\ref{sec:resultsMove} and \S\ref{sec:resultsBirth}, respectively; (i) the PDE cell mass in time compared to the mean number of cells across $10^3$ ABM simulations for different $N_\text{bir}$ values; and (j)--(l) the difference between the PDE and EA ABM solutions from (a)--(h). To provide more intuition, we overlay some cell positions from one ABM simulation. \label{fig:recMelFull}}
	\end{figure}

	One approach to addressing these discrepancies is to refit all three scaling parameters ($\aMM$, $\gamma$, and $c^+$) simultaneously, and we present the results of this approach in the SI-\S3. (Indeed, we show there that the errors produced with a simultaneous estimation approach can be relatively small, although the model parameters may not all be identifiable.) Because we are interested in understanding the interplay of individual-based mechanisms of proliferation and movement in continuous models, however, we instead take a simpler theoretical approach. Namely, we notice that the parameter $c^+$ is largely responsible for controlling the maximum cell density over long time periods. (We determine this by integrating Eqn.~\eqref{eqn:full1} over space and identifying the steady-state dynamics; this analysis reveals that equilibrium is reached when the density within any neighborhood $\Omega^\mathbf{x}_\mathrm{loc}$ is below $c^+$.) In order to limit the maximum density to our estimated empirical value of $400$ cells/mm$^2$ \cite{TakahashiMelDisperse,volkening2015}, we let $c^+ = 400|\Omega_\mathrm{loc}| \approx 7.0686$ cells. As we show in Fig.~\ref{fig:theoreticalC}(c), using this value of $c^+$, alongside our previously fit values of $\aMM$ and $\gamma$, produces PDE densities that are much closer to the corresponding ABM results. We thus fix $c^+ = 7.0686$ cells for the remainder of this manuscript, which allows us to highlight the time dynamics of our full PDE model in comparison to the EA ABM result with \textit{Box} and \textit{Offset rectangles} initial conditions in Fig.~\ref{fig:boxMelFull} and \ref{fig:recMelFull}, respectively.

	Fig.~\ref{fig:boxMelFull}(j), which depicts the time evolution of the estimated radius of support for the PDE and EA ABM results, shows that a reasonably accurate continuous description of the combined model can be obtained by using the scaling parameters obtained from a modular approach.  The supports of the ABM and PDE solutions both increase at roughly the same rate, although the degree to which the solutions agree can be affected by $N_\text{bir}$: when this parameter is small, the PDE solution travels at a faster rate than the ABM solution, whereas the opposite occurs when this value is large (greater than $N_\text{bir} = 50$ positions/day). At intermediate values of $N_\text{bir}$ (i.e., $N_\text{bir}=50$ positions/day), however, the ABM and PDE solution curves are almost identical. Fig. \ref{fig:boxMelFull}(i), which presents the number of cells over time, yields similar observations: the ABM and PDE solutions exhibit similar dynamics over the time period investigated here and there are certain values of $N_\text{bir}$ for which the solution curves are nearly identical. Fig. \ref{fig:recMelFull} further demonstrates that these observations do not depend on the specific choice of initial condition. Figs.~\ref{fig:boxMelFull} and \ref{fig:recMelFull} demonstrate that combining movement with proliferation also dissipates the oscillatory bands that we observed for movement alone in Fig.~\ref{fig:boxMel}. This is likely because the stochastic addition of cells in the birth model disrupts the regular cell spacing created by the movement model. Furthermore, the EA ABM and PDE solutions exhibit similar characteristic profiles without regions of high cell density around the edge of the initial condition support, in contrast to the birth-only model (Fig. \ref{fig:boxMelBirth}).

	\section{Discussion}\label{sec:discussion}

	We presented a procedure for constructing experimentally interpretable continuous models of cell migration and birth in biologically relevant settings of low numbers of individuals and localised interactions which may lie outside the validity of the mean-field regime. Specifically, we introduced and estimated scaling parameters in continuous models to account for realistic---i.e., relatively small and changing---numbers of cells with localised interactions. We applied this methodology to an illustrative, simplified example inspired by zebrafish pattern formation, in which we used a reduced ABM to generate individual-level data with biologically meaningful spatial and temporal units. Non-local rules for cell birth and migration, based on the ABM \cite{volkening2015}, informed our discrete and continuous descriptions and allowed us to transfer biological length scales and units to the macroscopic setting. Throughout our work, we stressed matching the spatio-temporal behaviour of our continuous and discrete models. We adopted a modular approach by estimating parameters in cases with either movement or birth before considering both mechanisms simultaneously. This allowed us to examine the specific contributions of each mechanism to self-organisation and provided insight into their interplay in discrete and continuous settings.
	
	We observed that the solutions of our continuous models expand at a different rate than EA ABM results and feature smoother profiles. Indeed, inaccuracies in mean-field descriptions for ``intermediate" numbers of individuals appear to be common in other biological phenomena described by simpler dynamics such as Fisher-KPP-type equations, cf. \cite{brunet1997, sontag2023}. In fact, both references analytically derive corrections to the wave speed, a procedure we cannot adopt due to our use of off-lattice models. However, this substantiates our introduction of scaling parameters to handle the discrepancy. By introducing and estimating parameters that rescale the time variable, we produced more accurate descriptions of agent-based movement or birth. However, when we used the same parameter values in a continuous model of both cell migration and birth in \S\ref{sec:resultsBoth}, the PDE did not produce close estimates of the full ABM. Specifically, our full continuous model yielded larger long-time densities than the EA ABM results, motivating us to re-estimate the threshold value $c^+$ with a theoretical approach. This generated a more faithful continuous description and highlighted that the effects of movement and proliferation are not simply additive. We thus stress that parameters must be fit to data in which all mechanisms of interest act simultaneously, in order to capture their interplay. This is particularly crucial for contexts such as cancer biology, where cell migration, proliferation, and death are known to play critical roles in tumour progression and immune response \cite{bull2022hallmarks}.

	Our results highlight how choices in numerical implementation affect parameter estimates and suggest several directions for future work that may improve our approach. For example, the optimal value of our parameter controlling the timescale of cell migration ($\aMM$) appears to be independent of the initial condition and the mesh resolution that we used to construct PDE solutions, provided the latter is sufficiently refined. One drawback of our current approach, however, is that we may need to estimate $\aMM$ and $\gamma$ for each new choice of discrete rules governing migration and cell birth, respectively, because these rules perturb the short-range interactions between relatively small numbers of cells. This naturally leads to the question of whether an analytic expression can be derived for these parameters. Several coarse-graining techniques that take into account higher-order correlations between cells in on- or off-lattice models may produce scaling factors similar to those introduced in this manuscript, but these may only apply to certain classes of ABMs \cite{bruna2017SIAMJAPPLMATH}. Alternatively, it may be possible to estimate scaling parameters of continuous models by investigating the convergence of EA ABM results to features of their solutions such as the speed of solution propagation, as in \cite{brunet1997,sontag2023} for on-lattice models; by accounting for the dynamics of the two-particle distribution, as in \cite{middleton2014continuum,binny2016collective}; or by exploring scaling relationships as in \cite{franke2022cell}. Adapting these approaches to our setting is an interesting avenue for future work.
 
    Additionally, our continuous models more accurately represent ABM results within the bulk of the solution support because the $L^2$ norm more strongly penalises discrepancies there. In the future, other norms, such as the $L^\infty$ error, could be used to match the solution supports given by our discrete and continuous models. Replacing our birth-rate scaling parameter $\gamma$ with a density-dependent function---either through rigorous derivation or an equation-learning approach \cite{NardiniLearning}---is another exciting future direction. In particular, because cell proliferation in the ABM involves selecting positions uniformly at random from the domain each day, the chance that we select a location $\textbf{z}$ that permits birth appears to depend in a nonlinear way on the solution support. More generally, our computational study does not provide theoretical explanations for our parameter values, and we plan to build on the intuition that we established here to develop these arguments in the future. 
	
	To simplify our initial study, we considered the dynamics of one cell population (i.e., melanophores in the main text and xanthophores in the SI-\S6), but pattern formation in zebrafish skin involves multiple cell types and longer-range interactions, as we highlight in Supplementary Fig.~1. Future work may extend our pipeline to construct more realistic continuous models with multiple cell types and interaction neighborhoods. Related to this, the initial conditions that we designed allowed us to make one-to-one comparisons between discrete- and continuous-model densities, but this may not always be possible. More realistic zebrafish models (i.e., \cite{volkening2015,volkening2018,volkening2020,Owen2020}) produce patterns that are more complicated than our box and stripe motifs. This means that ensemble-averaging stochastic ABM realisations may not retain information about the length scales inherent in patterns. For such cases, fitting parameters based on summary statistics (e.g., pair-correlation functions \cite{bull2022}, pattern-simplicity scores \cite{Miyazawa2020}, or persistent-homology approaches \cite{McGuirl2020}) may be more useful, and we plan to address this in future work. These and other directions move us toward constructing interpretable, analytically tractable continuous models of self-organisation, increasing our understanding of biological pattern formation more broadly.

	\appendix
	
	\section*{Code availability}
	Our model and parameter-fitting code is publicly available on GitHub \cite{repository}. Supplementary figures and videos are available on Figshare at \url{https://figshare.com/projects/Linking_discrete_and_continuous_models_of_cell_birth_and_migration_in_one_population/171234}. Data used to generate figures are given in Dryad \url{https://doi.org/10.5061/dryad.s4mw6m9cb}. 
	
	\section*{Acknowledgements}
	JAC and WDM were supported by the Advanced Grant Nonlocal-CPD (Nonlocal PDEs for Complex Particle Dynamics: Phase Transitions, Patterns and Synchronization) of the European Research Council Executive Agency (ERC) under the European Union’s Horizon 2020 research and innovation programme (grant agreement No. 883363). WDM, JAC, and AV would also like to thank the Isaac Newton Institute for Mathematical Sciences, Cambridge, for support and hospitality during the programme Mathematics of Movement, where work on this paper was undertaken. This work was supported by EPSRC grant no EP/R014604/1. CV acknowledges support from the Dr Perry James (Jim) Browne Research Centre on Mathematics and its Applications (University of Sussex). We are grateful to Shigeru Kondo for helpful discussions during an early stage of this research.
	
	\bibliography{references_v3.bib}

\begin{thebibliography}{10}

\bibitem{ME99}
Mogilner A, Edelstein-Keshet L.
\newblock A non-local model for a swarm.
\newblock J Math Biol. 1999;38(6):534-70.

\bibitem{d2006self}
D’Orsogna MR, Chuang YL, Bertozzi AL, Chayes LS.
\newblock Self-propelled particles with soft-core interactions: patterns,
  stability, and collapse.
\newblock Phys Rev Lett. 2006;96(10):104302.

\bibitem{cucker2007emergent}
Cucker F, Smale S.
\newblock Emergent behavior in flocks.
\newblock IEEE Trans Autom Control. 2007;52(5):852-62.

\bibitem{carrillo2010particle}
Carrillo JA, Fornasier M, Toscani G, Vecil F.
\newblock Particle, kinetic, and hydrodynamic models of swarming.
\newblock In: Mathematical modeling of collective behavior in socio-economic
  and life sciences. Springer; 2010. p. 297-336.

\bibitem{amack2012sorting}
Amack JD, Manning ML.
\newblock Knowing the boundaries: Extending the differential adhesion
  hypothesis in embryonic cell sorting.
\newblock Science. 2012;338:212-5.

\bibitem{burger2018sorting}
Burger M, Francesco MD, Fagioli S, Stevens A.
\newblock Sorting phenomena in a mathematical model for two mutually
  attracting/repelling species.
\newblock SIAM J Math Anal. 2018;50(3):3210-50.

\bibitem{carrillo2019population}
Carrillo JA, Murakawa H, Sato M, Togashi H, Trush O.
\newblock A population dynamics model of cell-cell adhesion incorporating
  population pressure and density saturation.
\newblock J Theor Biol. 2019;474:14-24.

\bibitem{Buttenschon2020}
Buttensch\"{o}n A, Edelstein-Keshet L.
\newblock Bridging from single to collective cell migration: A review of models
  and links to experiments.
\newblock PLOS Comput Biol. 2020;16(12).

\bibitem{tsai2022differentialadhesion}
Tsai TYC, Garner RM, Megason SG.
\newblock Adhesion-Based Self-Organization in Tissue Patterning.
\newblock Annu Rev Cell Dev Biol. 2022;38:349-74.

\bibitem{frohnhofer2013iridophores}
Frohnh{\"o}fer HG, Krauss J, Maischein HM, N{\"u}sslein-Volhard C.
\newblock Iridophores and their interactions with other chromatophores are
  required for stripe formation in zebrafish.
\newblock Development. 2013;140(14):2997-3007.

\bibitem{IrionRev2019}
Irion U, N\"{u}sslein-Volhard C.
\newblock The identification of genes involved in the evolution of color
  patterns in fish.
\newblock Curr Opin Genet Dev. 2019;57:31-8.

\bibitem{Parichy2021}
Parichy DM.
\newblock Evolution of pigment cells and patterns: recent insights from teleost
  fishes.
\newblock Curr Opin Genet Dev. 2021;69:88-96.

\bibitem{Kondo2021rev}
Kondo S, Watanabe M, Miyazawa S.
\newblock Studies of {T}uring pattern formation in zebrafish skin.
\newblock Philos Trans R Soc A. 2021;379(2213):20200274.

\bibitem{nakamasu2009interactions}
Nakamasu A, Takahashi G, Kanbe A, Kondo S.
\newblock Interactions between zebrafish pigment cells responsible for the
  generation of Turing patterns.
\newblock Proc Natl Acad Sci USA. 2009;106(21):8429-34.

\bibitem{singh2014proliferation}
Singh AP, Schach U, N{\"u}sslein-Volhard C.
\newblock Proliferation, dispersal and patterned aggregation of iridophores in
  the skin prefigure striped colouration of zebrafish.
\newblock Nat Cell Biol. 2014;16(6):604-11.

\bibitem{Gur2020}
Gur D, Bain EJ, Johnson KR, Aman AJ, Amalia~Pasolli H, Flynn JD, et~al.
\newblock In situ differentiation of iridophore crystallotypes underlies
  zebrafish stripe patterning.
\newblock Nat Commun. 2020;11(1):6391.

\bibitem{hamada2014involvement}
Hamada H, Watanabe M, Lau HE, Nishida T, Hasegawa T, Parichy DM, Kondo S.
\newblock Involvement of Delta/Notch signaling in zebrafish adult pigment
  stripe patterning.
\newblock Development. 2014;141(2):318-24.

\bibitem{Inaba}
Inaba M, Yamanaka H, Kondo S.
\newblock Pigment pattern formation by contact-dependent depolarization.
\newblock Science. 2012;335(6069):677-7.

\bibitem{Yamaguchi_Yoshimoto_Kondo_2007}
Yamaguchi M, Yoshimoto E, Kondo S.
\newblock Pattern regulation in the stripe of zebrafish suggests an underlying
  dynamic and autonomous mechanism.
\newblock Proc Natl Acad Sci USA. 2007;104(12):4790–4793.

\bibitem{kroll2021simple}
Kroll F, Powell GT, Ghosh M, Gestri G, Antinucci P, Hearn TJ, et~al.
\newblock A simple and effective F0 knockout method for rapid screening of
  behaviour and other complex phenotypes.
\newblock eLife. 2021;10:e59683.

\bibitem{volkening2015}
Volkening A, Sandstede B.
\newblock Modelling stripe formation in zebrafish: an agent-based approach.
\newblock J R Soc Interface. 2015;12(112):20150812.

\bibitem{fadeev2015}
Fadeev A, Krauss J, Frohnhöfer HG, Irion U, Nüsslein-Volhard C.
\newblock Tight Junction Protein 1a regulates pigment cell organisation during
  zebrafish colour patterning.
\newblock eLife. 2015;4:e06545.

\bibitem{byrne2010}
Byrne HM.
\newblock Dissecting cancer through mathematics: From the cell to the animal
  model.
\newblock Nat Rev Cancer. 2010;10:221-30.

\bibitem{gompper2020}
Gompper G, Winkler RG, Speck T, Solon A, Nardini C, Peruani F, et~al.
\newblock The 2020 motile active matter roadmap.
\newblock J Phys Condens Matter. 2020;32:193001.

\bibitem{stillman2023}
Stillman NR, Mayor R.
\newblock Generative models of morphogenesis in developmental biology.
\newblock Semin Cell Dev Biol. 2023;147:83-90.

\bibitem{VolkeningRev}
Volkening A.
\newblock Linking genotype, cell behavior, and phenotype: multidisciplinary
  perspectives with a basis in zebrafish patterns.
\newblock Curr Opin Genet Dev. 2020;63:78-85.

\bibitem{alert2020}
Alert R, Trepat X.
\newblock Physical models of collective cell migration.
\newblock Annu Rev Condens Matter Phys. 2020;11:77-101.

\bibitem{metzcar2019}
Metzcar J, Wang Y, Heiland R, Macklin P.
\newblock A review of cell-based computational modeling in cancer biology.
\newblock JCO Clin Cancer Inform. 2019;2:1-13.

\bibitem{deutsch2015cellular}
Deutsch A.
\newblock Cellular automaton models for collective cell behaviour.
\newblock In: Cellular Automata and Discrete Complex Systems. AUTOMATA 2015.
  Lecture Notes in Computer Science. Springer; 2015. p. 1-10.

\bibitem{CellularAutomaton17}
Deutsch A, Dormann S.
\newblock Cellular Automaton Modeling of Biological Pattern Formation,
  Characterization, Examples, and Analysis.
\newblock Springer; 2017.

\bibitem{hirashima2017}
Hirashima T, Rens EG, Merks RMH.
\newblock Cellular Potts modeling of complex multicellular behaviors in tissue
  morphogenesis.
\newblock Dev Growth Differ. 2017;59(5):329-39.

\bibitem{rens2019}
Rens EG, Edelstein-Keshet L.
\newblock From energy to cellular forces in the Cellular Potts Model: An
  algorithmic approach.
\newblock PLOS Comput Biol. 2019;15(12):1-23.

\bibitem{alt2017vertex}
Alt S, Ganguly P, Salbreux G.
\newblock Vertex models: from cell mechanics to tissue morphogenesis.
\newblock Philos Trans R Soc B. 2017;372(1720):20150520.

\bibitem{fletcher2014}
Fletcher A, Osterfield M, Baker R, Shvartsman S.
\newblock Vertex Models of Epithelial Morphogenesis.
\newblock Biophys J. 2014;106(11):2291-304.

\bibitem{Bullara}
Bullara D, De~Decker Y.
\newblock Pigment cell movement is not required for generation of {T}uring
  patterns in zebrafish skin.
\newblock Nat Commun. 2015;6(6971).

\bibitem{MorDeutsch}
Moreira J, Deutsch A.
\newblock Pigment pattern formation in zebrafish during late larval stages: A
  model based on local interactions.
\newblock Dev Dyn. 2005;232(1):33-42.

\bibitem{Owen2020}
Owen JP, Kelsh RN, Yates CA.
\newblock A quantitative modelling approach to zebrafish pigment pattern
  formation.
\newblock eLife. 2020;9:e52998.

\bibitem{caicedo2008silico}
Caicedo-Carvajal CE, Shinbrot T.
\newblock In silico zebrafish pattern formation.
\newblock Dev Biol. 2008;315(2):397-403.

\bibitem{volkening2020}
Volkening A, Abbott MR, Chandra N, Dubois B, Lim F, Sexton D, Sandstede B.
\newblock Modeling stripe formation on growing zebrafish tailfins.
\newblock Bull Math Biol. 2020;82(56).

\bibitem{volkening2018}
Volkening A, Sandstede B.
\newblock Iridophores as a source of robustness in zebrafish stripes and
  variability in \textit{{D}anio} patterns.
\newblock Nat Commun. 2018;9(3231).

\bibitem{osborne2017}
Osborne JM, Fletcher AG, Pitt-Francis JM, Maini PK, Gavaghan DJ.
\newblock Comparing individual-based approaches to modelling the
  self-organization of multicellular tissues.
\newblock PLOS Comput Biol. 2017;13:e1005387.

\bibitem{murray2002mathematical}
Murray JD.
\newblock Mathematical biology: I. An introduction.
\newblock Springer; 2002.

\bibitem{perthame2006transport}
Perthame B.
\newblock Transport equations in biology.
\newblock Springer Science \& Business Media; 2006.

\bibitem{Turing}
Turing AM.
\newblock The chemical basis of morphogenesis.
\newblock Philos Trans R Soc Lond B. 1952;237(641):37-72.

\bibitem{maini1997spatial}
Maini P, Painter K, Chau HP.
\newblock Spatial pattern formation in chemical and biological systems.
\newblock J Chem Soc. 1997;93(20):3601-10.

\bibitem{marciniak2017instability}
Marciniak-Czochra A, Karch G, Suzuki K.
\newblock Instability of Turing patterns in reaction-diffusion-ODE systems.
\newblock J Math Biol. 2017;74:583-618.

\bibitem{armstrong2006continuum}
Armstrong NJ, Painter KJ, Sherratt JA.
\newblock A continuum approach to modelling cell--cell adhesion.
\newblock J Theor Biol. 2006;243(1):98-113.

\bibitem{murakawa2015continuous}
Murakawa H, Togashi H.
\newblock Continuous models for cell--cell adhesion.
\newblock J Theor Biol. 2015;374:1-12.

\bibitem{carrillo2018zoology}
Carrillo JA, Huang Y, Schmidtchen M.
\newblock Zoology of a nonlocal cross-diffusion model for two species.
\newblock SIAM J Appl Math. 2018;78(2):1078-104.

\bibitem{carrillo2018adhesion}
Carrillo JA, Colombi A, Scianna M.
\newblock Adhesion and volume constraints via nonlocal interactions determine
  cell organisation and migration profiles.
\newblock J Theor Biol. 2018;445:75-91.

\bibitem{Gaffney}
Gaffney EA, Seirin~Lee S.
\newblock {The sensitivity of Turing self-organization to biological feedback
  delays: 2D models of fish pigmentation}.
\newblock Math Med Biol. 2015;32:57-79.

\bibitem{Konow2021}
Konow C, Li Z, Shepherd S, Bullara D, Epstein IR.
\newblock Influence of survival, promotion, and growth on pattern formation in
  zebrafish skin.
\newblock Sci Rep. 2021;11(9864).

\bibitem{Bloomfield}
Bloomfield JM, Painter KJ, Sherratt JA.
\newblock How does cellular contact affect differentiation mediated pattern
  formation?
\newblock Bull Math Biol. 2011;73(7):1529-58.

\bibitem{kondo2017updated}
Kondo S.
\newblock An updated kernel-based Turing model for studying the mechanisms of
  biological pattern formation.
\newblock J Theor Biol. 2017;414:120-7.

\bibitem{painter}
Painter KJ, Bloomfield JM, Sherratt JA, Gerisch A.
\newblock A nonlocal model for contact attraction and repulsion in
  heterogeneous cell populations.
\newblock Bull Math Biol. 2015;77(6):1132-65.

\bibitem{giacomin1997phase}
Giacomin G, Lebowitz JL.
\newblock Phase segregation dynamics in particle systems with long range
  interactions. I. Macroscopic limits.
\newblock J Stat Phys. 1997;87:37-61.

\bibitem{painter2002volume}
Painter KJ, Hillen T.
\newblock Volume-filling and quorum-sensing in models for chemosensitive
  movement.
\newblock Can Appl Math Quart. 2002;10(4):501-43.

\bibitem{penington2011building}
Penington CJ, Hughes BD, Landman KA.
\newblock Building macroscale models from microscale probabilistic models: a
  general probabilistic approach for nonlinear diffusion and multispecies
  phenomena.
\newblock Physical Review E. 2011;84(4):041120.

\bibitem{hillen2013}
Hillen T, Painter KJ.
\newblock Transport and anisotropic diffusion models for movement in oriented
  habitats.
\newblock Lect Notes Math. 2013;2071:177-222.

\bibitem{carrillo2014derivation}
Carrillo JA, Choi YP, Hauray M.
\newblock The derivation of swarming models: Mean-field limit and Wasserstein
  distances.
\newblock In: Muntean A, Toschi F, editors. Collective Dynamics from Bacteria
  to Crowds: An Excursion Through Modeling, Analysis and Simulation. CISM
  International Centre for Mechanical Sciences. Vienna: Springer Vienna; 2014.
  p. 1-46.

\bibitem{CK14}
Chen Y, Kolokolnikov T.
\newblock A minimal model of predator–swarm interactions.
\newblock J R Soc Interface. 2014;11:20131208.

\bibitem{bruna22}
Bruna M, Burger M, Pietschmann JF, Wolfram MT.
\newblock Active crowds.
\newblock In: Active particles, {V}ol. 3. Model. Simul. Sci. Eng. Technol..
  Birkh\"{a}user/Springer, Cham; 2022. p. 35-73.

\bibitem{champagnat2006unifying}
Champagnat N, Ferri{\`e}re R, M{\'e}l{\'e}ard S.
\newblock Unifying evolutionary dynamics: from individual stochastic processes
  to macroscopic models.
\newblock Theor Popul Biol. 2006;69(3):297-321.

\bibitem{champagnat2008individual}
Champagnat N, Ferri{\`e}re R, M{\'e}l{\'e}ard S.
\newblock Individual-based probabilistic models of adaptive evolution and
  various scaling approximations.
\newblock Prog Probab. 2008;59:75.

\bibitem{champagnat2008individual2}
Champagnat N, Ferri{\`e}re R, M{\'e}l{\'e}ard S.
\newblock From individual stochastic processes to macroscopic models in
  adaptive evolution.
\newblock Stoch Models. 2008;24:2-44.

\bibitem{hansen2006mcdonaldi}
Hansen JP, McDonald IR.
\newblock Theory of Simple Liquids.
\newblock London: Academic Press; 2006.

\bibitem{lushnikov2008}
Lushnikov PM, Chen N, Alber M.
\newblock Macroscopic dynamics of biological cells interacting via chemotaxis
  and direct contact.
\newblock Phys Rev E. 2008;78:061904.

\bibitem{simpson2009exclusion}
Simpson MJ, Landman KA, Hughes BD.
\newblock Multi-species simple exclusion processes.
\newblock Phys A: Stat Mech. 2009;388:399-406.

\bibitem{simpson2011meanfield}
Simpson MJ, Baker RE.
\newblock Corrected mean-field models for spatially dependent
  advection-diffusion-reaction phenomena.
\newblock Phys Rev E. 2011;83:51922.

\bibitem{markham2013}
Markham DC, Simpson MJ, Baker RE.
\newblock Simplified method for including spatial correlations in mean-field
  approximations.
\newblock Phys Rev E. 2013;87:62702.

\bibitem{wieczorek2023}
Wieczorek R.
\newblock Hydrodynamic limit of a stochastic model of proliferating cells with
  chemotaxis.
\newblock Kinet Relat Models. 2023;16:373-93.

\bibitem{binny2015spatial}
Binny RN, Plank MJ, James A.
\newblock Spatial moment dynamics for collective cell movement incorporating a
  neighbour-dependent directional bias.
\newblock Journal of the Royal Society Interface. 2015;12(106):20150228.

\bibitem{binny2016collective}
Binny RN, James A, Plank MJ.
\newblock Collective cell behaviour with neighbour-dependent proliferation,
  death and directional bias.
\newblock Bulletin of Mathematical Biology. 2016;78:2277-301.

\bibitem{middleton2014continuum}
Middleton AM, Fleck C, Grima R.
\newblock A continuum approximation to an off-lattice individual-cell based
  model of cell migration and adhesion.
\newblock Journal of theoretical biology. 2014;359:220-32.

\bibitem{bruna2017diffusion}
Bruna M, Chapman SJ, Robinson M.
\newblock Diffusion of particles with short-range interactions.
\newblock SIAM Journal on Applied Mathematics. 2017;77(6):2294-316.

\bibitem{johnston2017new}
Johnston ST, Baker RE, Simpson MJ.
\newblock A new and accurate continuum description of moving fronts.
\newblock New Journal of Physics. 2017;19(3):033010.

\bibitem{bruna2012excluded}
Bruna M, Chapman SJ.
\newblock Excluded-volume effects in the diffusion of hard spheres.
\newblock Physical Review E. 2012;85(1):011103.

\bibitem{Patterson2014}
Patterson LB, Bain EJ, Parichy DM.
\newblock Pigment cell interactions and differential xanthophore recruitment
  underlying zebrafish stripe reiteration and \emph{{D}anio} pattern evolution.
\newblock Nat Commun. 2014;5(5299).

\bibitem{Walderich2016}
Walderich B, Singh AP, Mahalwar P, N\"{u}sslein-Volhard C.
\newblock Homotypic cell competition regulates proliferation and tiling of
  zebrafish pigment cells during colour pattern formation.
\newblock Nat Commun. 2016;7(1):11462.

\bibitem{golse}
Golse F.
\newblock The mean-field limit for the dynamics of large particle systems.
\newblock Journ\'ees \'Equations aux d\'eriv\'ees Partielles. 2003.

\bibitem{DF2013}
Di~Francesco M, Fagioli S.
\newblock Measure solutions for non-local interaction PDEs with two species.
\newblock Nonlinearity. 2013;26(10):2777.

\bibitem{TakahashiMelDisperse}
Takahashi G, Kondo S.
\newblock Melanophores in the stripes of adult zebrafish do not have the nature
  to gather, but disperse when they have the space to move.
\newblock Pigment Cell Melanoma Res. 2008;21(6):677-86.

\bibitem{ParTur256}
Parichy DM, Turner JM.
\newblock Zebrafish \emph{puma} mutant decouples pigment pattern and somatic
  metamorphosis.
\newblock Dev Biol. 2003;256(2):242-57.

\bibitem{chiari2022hybrid}
Chiari G, Delitala ME, Morselli D, Scianna M.
\newblock A hybrid modeling environment to describe aggregates of cells
  heterogeneous for genotype and behavior with possible phenotypic transitions.
\newblock International Journal of Non-Linear Mechanics. 2022;144:104063.

\bibitem{brunet1997}
Brunet E, Derrida B.
\newblock Shift in the velocity of a front due to a cutoff.
\newblock Phys Rev E. 1997;56:2597-604.

\bibitem{sontag2023}
Sontag A, Rogers T, Yates CA.
\newblock Stochastic drift in discrete waves of nonlocally interacting
  particles.
\newblock Phys Rev E. 2023;107:014128.

\bibitem{bull2022hallmarks}
Bull JA, Byrne HM.
\newblock The hallmarks of mathematical oncology.
\newblock Proc IEEE. 2022;110(5):523-40.

\bibitem{bruna2017SIAMJAPPLMATH}
Bruna M, Chapman SJ, Robinson M.
\newblock Diffusion of Particles with Short-Range Interactions.
\newblock SIAM J Appl Math. 2017;77(6):2294-316.
\newblock Available from: \url{https://doi.org/10.1137/17M1118543}.

\bibitem{franke2022cell}
Franke F, Aland S, B{\"o}hme HJ, Voss-B{\"o}hme A, Lange S.
\newblock Is cell segregation like oil and water: asymptotic versus transitory
  regime.
\newblock PLoS Comput Biol. 2022;18(9):e1010460.

\bibitem{NardiniLearning}
Nardini JT, Baker RE, Simpson MJ, Flores KB.
\newblock Learning differential equation models from stochastic agent-based
  model simulations.
\newblock J Roy Soc Interface. 2021;18:20200987.

\bibitem{bull2022}
Bull JA, Byrne HM.
\newblock Quantification of spatial and phenotypic heterogeneity in an
  agent-based model of tumour-macrophage interactions.
\newblock PLOS Comput Biol. 2023;19(3):e1010994.

\bibitem{Miyazawa2020}
Miyazawa S.
\newblock Pattern blending enriches the diversity of animal colorations.
\newblock Sci Adv. 2020;6(49).

\bibitem{McGuirl2020}
McGuirl MR, Volkening A, Sandstede B.
\newblock Topological data analysis of zebrafish patterns.
\newblock Proc Natl Acad Sci USA. 2020;117(10):5113-24.

\bibitem{repository}
Martinson WD, Schmitchen M, Volkening A, Venkataraman C, Carrillo JA. GitHub
  repository for ``Linking discrete and continuous models of cell birth and
  migration". GitHub; 2023.
\newblock \url{https://github.com/wdmartinson/Self-Organization-One-Species}.

\end{thebibliography}
	\bibliographystyle{Vancouver}
	
\end{document}